\documentclass[aps,prd,preprintnumbers,groupedaddress,nofootinbib,amssymb,eqsecnum,notitlepage,twocolumn,epsfig]{revtex4-1}

\usepackage{graphicx}
\usepackage{graphicx}
\usepackage{bm}
\usepackage{amsmath,amsthm,amssymb}
\usepackage{enumitem}
\usepackage{color}
\definecolor{ultramarine}{rgb}{0.07, 0.04, 0.56}
\definecolor{cadmiumgreen}{rgb}{0.0, 0.42, 0.24}
\definecolor{indigo(dye)}{rgb}{0.0, 0.25, 0.42}
\definecolor{orangered}{RGB}{255,69,0}
\usepackage[linktocpage=true]{hyperref}
\hypersetup{
colorlinks=true,
citecolor=ultramarine,
linkcolor=cadmiumgreen,
urlcolor=indigo(dye),
}

\begin{document}
\newcommand{\newc}{\newcommand}

\newc{\be}{\begin{equation}}
\newc{\ee}{\end{equation}}
\newc{\ba}{\begin{eqnarray}}
\newc{\ea}{\end{eqnarray}}
\newc{\bea}{\begin{eqnarray*}}
\newc{\eea}{\end{eqnarray*}}
\newc{\D}{\partial}
\newc{\ie}{{\it \emph{i.e.}} }
\newc{\eg}{{\it e.g.} }
\newc{\etc}{{\it etc.} }
\newc{\etal}{{\it et al.}}
\newcommand{\nn}{\nonumber}
\newc{\ra}{\rightarrow}
\newc{\lra}{\leftrightarrow}
\newc{\lsim}{\buildrel{<}\over{\sim}}
\newc{\gsim}{\buildrel{>}\over{\sim}}
\def\mpl{M_{\rm pl}}
\def\d{\mathrm{d}}

\newcommand{\he}[1]{\textcolor{red}{#1}}

\title{Inflation with mixed helicities and its observational imprint on CMB}

\author{
Lavinia Heisenberg$^{1}$, H\'ector Ram\'irez$^{2,3}$, and  
Shinji Tsujikawa$^{4}$}

\affiliation{
$^1$Institute for Theoretical Physics, ETH Zurich, Wolfgang-Pauli-Strasse 27, 8093 Zurich, Switzerland\\
$^2$Departamento de F\'{i}sica Te\'orica, Universidad de Valencia,
Dr. Moliner 50, E-46100 Burjassot, Spain\\
$^3$Instituto de F\'{i}sica Corpuscular (IFIC), Universidad de Valencia-CSIC,
E-46980, Paterna, Spain\\
$^4$Department of Physics, Faculty of Science, 
Tokyo University of Science,
1-3, Kagurazaka, Shinjuku-ku, Tokyo 162-8601, Japan}

\date{\today}

\begin{abstract}

In the framework of effective field theories with prominent helicity-0 and 
helicity-1 fields coupled to each other via a dimension-3 operator, 
we study the dynamics of inflation driven by the helicity-0 mode, with 
a given potential energy, as well as the evolution of cosmological perturbations, 
influenced by the presence of a mixing term between both helicities. 
In this scenario, the temporal component of the helicity-1 mode is 
an auxiliary field and can be integrated out in terms of the time 
derivative of the helicity-0 mode, so that the background dynamics 
effectively reduces to that in single-field inflation modulated 
by a parameter $\beta$ associated to the coupling between 
helicity-0 and helicity-1 modes.
We discuss the evolution of a longitudinal scalar perturbation $\psi$ 
and an inflaton fluctuation $\delta \phi$, and 
explicitly show that a particular combination of these two,
which corresponds to an isocurvature mode, 
is subject to exponential suppression by the vector mass comparable 
to the Hubble expansion rate during inflation.
Furthermore, we find that the effective single-field description 
corrected by $\beta$ also holds for 
the power spectrum of curvature perturbations generated 
during inflation. We compute the standard inflationary observables such as 
the scalar spectral index $n_s$ and the tensor-to-scalar ratio $r$ 
and confront several inflaton potentials with the 
recent observational data provided by \emph{Planck} 2018. 
Our results show that the coupling between helicity-0 and 
helicity-1 modes can lead to a smaller value of the
tensor-to-scalar ratio especially for small-field inflationary models, 
so our scenario exhibits even better compatibility with the current  
observational data.

\end{abstract}


\maketitle

\section{Introduction}
\label{introsec}

Inflation \cite{Sta80,oldinf} provides a causal mechanism for generating 
primordial density perturbations responsible for large-scale structures 
of the Universe \cite{oldper}. 
Moreover, the temperature anisotropies observed in the
Cosmic Microwave Background (CMB) are overall consistent with 
the prediction of the inflationary paradigm \cite{WMAP,Planck2015,Planck2018}. 
It is anticipated that the possible detection of B-mode polarizations 
in the future will offer the opportunity to identify the origin of inflation.
 
The simplest candidate for inflation is 
a new scalar field $\phi$ beyond the Standard Model subject to 
a particular potential $V(\phi)$. 
As long as the field evolves slowly along a nearly flat potential, 
the primordial power spectra of scalar and tensor perturbations 
generated during inflation are close to 
scale-invariant \cite{Kolb}. 
The deviation from scale invariance, characterized by the spectral index $n_s$ and the 
tensor-to-scalar ratio $r$, depends strongly on the assumption about the 
inflaton potential. Using the bounds of $n_s$ and $r$ 
constrained from the CMB data, one can distinguish between 
different inflationary 
models \cite{Martin,TOKA,PTEP,Planck2015,Planck2018,Escudero:2015wba}. 

A cosmological accelerated expansion can be driven not only by 
a scalar field but also by a vector field.
Indeed, the accelerated solutions were found in Ref.~\cite{Koi08,Maroto1} 
in traditional vector-tensor theories, however they are generically 
plagued by instabilities \cite{vecins1,vecins2,vecins3}. 
In the so-called generalized Proca theories 
where an abelian vector field with broken $U(1)$ gauge symmetry has 
derivative self-interactions and nonminimal couplings 
to gravity \cite{Heisenberg,Allys,Jimenez16} (see also Ref.~\cite{BGP}), 
the existence of a temporal vector component $A_0$ can give rise to 
de Sitter solutions. 
Indeed, the generalized Proca theories are very successful 
for describing the late-time cosmic 
acceleration \cite{GPcosmo1,GPcosmo2}. 

On the other hand, there are also mechanisms for realizing the cosmic acceleration 
by using space-like vector fields \cite{Bento,Armen}. 
Naively this configuration is not 
compatible with an isotropic cosmological background, but 
the rotational invariance can be preserved by considering 
three orthogonal vector fields aligned with three spatial directions.
Indeed, three vector fields $A^a_{\mu}$ nonminimally coupled 
to the Ricci scalar $R$ in the form $RA^a_{\mu}A^{a\mu}$ 
can lead to inflation \cite{vectorinf}, but such accelerated solutions 
are plagued by either ghosts or Laplacian instabilities \cite{Peloso}. 
Non-abelian gauge fields with $SU(2)$ gauge symmetry can 
be also the source for inflation without instabilities \cite{gaugeinf,Soda},
but the scalar spectral index $n_s$ and the tensor-to-scalar ratio $r$ 
are not compatible with the CMB data \cite{Namba,Ads}. There exists an 
inflationary scenario driven by a nonminimally coupled 
non-abelian gauge field \cite{Davydov}, but the tensor perturbation is 
subject to ghost instabilities \cite{Jose}.

Efforts have also been made to construct well-behaved inflationary models 
in the presence of vector fields but where, as in the standard case, the main 
source for the accelerated expansion is a scalar field $\phi$. 
It is of particular interest the case where this field is coupled to 
an abelian vector field $A_{\mu}$. 
It is known that, for this type of scenarios,
a stable inflationary solution with an anisotropic 
hair exists for the coupling of the form 
$f^2 (\phi) F_{\mu \nu}F^{\mu \nu}$, 
where $F_{\mu \nu}=\nabla_{\mu}A_{\nu}
-\nabla_{\nu}A_{\mu}$ is the field strength tensor with a covariant derivative 
operator $\nabla_{\mu}$ \cite{aniinf}. 
The same coupling has been often 
used for the generation of magnetic fields during 
inflation \cite{Turner,Ratra}, 
but in such cases the models need to be carefully constructed 
to avoid the back-reaction and 
strong-coupling problems \cite{Bamba,Kanno,Slava,Fujita,Mukoh}.

Moreover, in the presence of a real scalar field $\phi$ and a vector field $A_{\mu}$ with derivative 
self-interactions and nonminimal couplings to gravity, 
the general action of scalar-vector-tensor (SVT) theories 
was recently constructed by keeping the equations of motion 
up to second order \cite{Heisenberg2}. 
In particular, the massive vector field with broken $U(1)$ 
gauge symmetry is relevant to the cosmological application. 
In this case, the vector perturbation is subject to 
exponential suppression by the mass of $A_{\mu}$.

Among the possible interactions between 
scalar and massive vector fields, and in particular for inflation, 
the coupling $A^{\mu} \nabla_{\mu} \phi$ is the simplest one 
modifying the inflaton velocity, $\dot{\phi}$,
during the cosmic expansion. This interaction is not only prone to 
SVT theories but arises in many effective field theories as one of the
lowest-order operator, once the involved broken gauge symmetries
are compensated by the introduction of appropriate St\"uckelberg fields.
In addition, the vector-field contribution to the 
total energy density during inflation is subdominant 
relative to the scalar potential $V(\phi)$, yet 
the modification to the inflaton velocity induced by the vector field 
can affect the primordial power spectra of scalar 
and tensor perturbations. 
See Ref.~\cite{Heisenberg:2018vsk} for a recent review on the 
systematic construction of modified gravity theories based on 
additional scalar, vector and tensor fields (see also \cite{Amendola:2012ys}).

For the aforementioned type of interaction, $A^\mu\nabla_\mu\phi$, 
there exists a longitudinal scalar perturbation, $\psi$, arising from $A_\mu$, 
besides the inflaton fluctuation $\delta\phi$~\cite{HKT18a,KT18,HKT18b}. 
This longitudinal perturbation contributes to the total curvature perturbation 
$\mathcal{R}$ in a nontrivial way. 
Therefore, the computation of the primordial power spectrum, incorporating 
both $\psi$ and $\delta\phi$, is not as straightforward as in the standard 
canonical case. In this paper, we address this problem and derive the standard inflationary 
observables such as $n_s$ and $r$ under the slow-roll approximation. 
We show that, as in the canonical case, one can relate these observables with 
slow-roll parameters but with a rescaling factor $\beta$ coming from the helicity-0 
and helicity-1 mixing. Using these general expressions, 
we then confront several different inflaton 
potentials with the recent CMB data  
provided by the 2018 results from the \emph{Planck} collaboration~\cite{Planck2018}.

This paper is organized as follows. 
In Sec.~\ref{modelsec}, we discuss the background inflationary 
dynamics and show that the system effectively reduces to 
that of a  single-field inflation. 
In Sec.~\ref{tvsec}, we revisit the primordial tensor power 
spectrum generated in our scenario and also study 
the evolution of vector perturbations during inflation.
In Sec.~\ref{persec}, we investigate how the perturbations $\psi$ 
and $\delta \phi$ evolve during inflation and obtain the resulting 
power spectrum of total curvature perturbations.
In Sec.~\ref{obsersec}, we compute inflationary observables 
and test several inflaton potentials with the latest 
\emph{Planck} 2018 data.
Sec.~\ref{consec} is devoted to conclusions.

\section{Inflation with a scalar-vector coupling}
\label{modelsec}

In many effective field theories, mixings between different helicity modes, 
even with derivative interactions, arise in a natural way. 
In massive gravity and massive Proca theories, the decomposition 
of helicities yields interesting couplings among them \cite{deRham:2012ew,Heisenberg,Jimenez16} ---this, in fact, motivated 
the construction of SVT theories \cite{Heisenberg2}.
The particular mixing of the form $A^{\mu}\nabla_{\mu} \phi$ 
arises quite naturally 
and is a unique coupling that modifies the involved propagators of 
scalar and vector fields. 
As we will see below, one possible origin of this coupling is the 
standard Proca mass term, which modifies the property of 
propagator by the mass parameter.

Let us consider, for instance, the Lagrangian of 
the standard Proca field:
\be
\label{LagProca}
\mathcal L_{A_\mu}=-\frac14F_{\mu\nu}F^{\mu\nu}
-\frac12M^2A_\mu A^\mu\,.
\ee
The existence of the mass term $M$ explicitly breaks 
the $U(1)$ gauge symmetry and therefore the massive 
spin-1 field propagates 3 degrees of freedom. 
Since the gauge invariance is just a redundancy, 
one can restore it by introducing
a St\"uckelberg field $\phi$ via the field transformation
\be
A_\mu \to A_\mu + \nabla_\mu \phi.
\ee
The initial Lagrangian for the massive spin-1 field (\ref{LagProca}) 
then modifies to
\be
\mathcal L_{A_\mu}=-\frac14F_{\mu\nu}F^{\mu \nu}
-\frac12M^2(A_\mu+\nabla_\mu\phi)(A^\mu+\nabla^\mu\phi)\,. 
\ee
Notice that the kinetic term $-F_{\mu\nu}F^{\mu \nu}/4$
is not modified under this change of variables since it is gauge invariant. 
Here, the helicity-0 field $\phi$ represents the longitudinal mode of 
the massive vector field. 
Written in this form, the standard Proca theory is now invariant under 
the simultaneous transformations $A_\mu \to A_\mu + \nabla_\mu \theta$ 
and $\phi \to \phi-\theta$. After canonically normalizing the St\"uckelberg field 
$\phi\to \phi/M$, the Lagrangian becomes 
\ba
\mathcal L_{A_\mu}
&=&-\frac14F_{\mu\nu}F^{\mu \nu}
-\frac12M^2 A_\mu A^{\mu}
-\frac12\nabla_{\mu}\phi \nabla^{\mu}\phi \nonumber \\
& &
-MA^\mu \nabla_\mu\phi\,.
\label{eq:lagprocafin}
\ea
The last term is exactly the coupling we are interested in. 
This Lagrangian constitutes our low energy effective field theory. 

In the following, we will consider a soft breaking of the shift symmetry 
of the helicity-0 mode and introduce a scalar potential $V(\phi)$ 
of the real scalar field $\phi$ for the purpose of realizing 
a successful inflationary 
scenario. Bear in mind that any UV completion will unavoidably 
introduce the breaking of global symmetry anyway.
Our setup consists in an inflationary scenario
in which the inflaton field $\phi$ has a derivative interaction 
with a massive vector field $A^{\mu}$ of the form 
$A^{\mu}\nabla_{\mu} \phi$, 
equivalent to that in Eq.~\eqref{eq:lagprocafin}.
The inflationary period is mostly driven by the scalar potential 
$V(\phi)$, but the scalar-vector coupling modifies the dynamics 
of inflation and the primordial power spectra of cosmological 
perturbations. We then focus on the action~\footnote{It is worth emphasizing that this model propagates six degrees of freedom: 2 as in standard GR, 3 from the massive vector field and 1 from the scalar field. The Proca Lagrangian in \eqref{LagProca} written as \eqref{eq:lagprocafin}, on the other hand, propagates only five degrees of freedom (including gravity). After introducing the St\"uckelberg field, the Proca vector field becomes gauge invariant and the longitudinal mode of the initial Proca field is transformed into the St\"uckelberg field itself. By including a general potential term for the scalar field, we explicitly break the previously restored gauge symmetry (or the related shift symmetry of the scalar field) and the theory propagates one more degree of freedom. This serves just as illustrative purposes, namely that the operator $A^{\mu}\nabla_{\mu} \phi$ is a hermitian operator.}

\ba
{\cal S}&=&\int d^4 x \sqrt{-g} 
\Biggl[ \frac{M_{\rm pl}^2}{2}R
+F+X_1-V(\phi) \nonumber \\
& &\qquad \qquad \quad~
+\beta_m M X_2+\beta_A M^2 X_3 \Biggr]\,,
\label{action}
\ea
where $g$ is the determinant of a metric tensor $g_{\mu \nu}$, 
$M_{\rm pl}$ is the reduced Planck mass, $R$ 
is the Ricci scalar, and $F=-(1/4)F_{\mu \nu}F^{\mu \nu}$.
The quantity $X_1$ is the scalar kinetic energy 
$X_1=-(1/2) \nabla_{\mu} \phi \nabla^{\mu} \phi$, while 
$X_2$ and $X_3$ are defined by 
\be
X_2=-\frac{1}{2} A^{\mu}\nabla_{\mu} \phi,\qquad 
X_3=-\frac{1}{2} A_{\mu}A^{\mu}\,.
\ee
In the last two terms of Eq.~(\ref{action}), $M$ is a positive 
constant (mass of the vector field) relevant to the mass scale of inflation, 
and $\beta_m$ and $\beta_A$ are 
dimensionless constants associated with the 
scalar-vector mixing and the vector mass, respectively. 

To discuss the background dynamics of inflation, 
we consider the flat Friedmann-Lema\^{i}tre-Robertson-Walker (FLRW) 
spacetime described by the line element $ds^2=-dt^2+a^2(t) \delta_{ij}dx^i dx^j$, 
where $a(t)$ is a time-dependent scale factor. 
The vector-field profile compatible with this metric
is of the form $A_{\mu}=(A_0(t), 0, 0, 0)$, 
with a time-dependent scalar field $\phi=\phi(t)$.
The background equations of motion in full parity-invariant 
SVT theories were already derived in Refs.~\cite{HKT18a,KT18}.
For the action (\ref{action}), they are given by 
\ba
& &
3M_{\rm pl}^2 H^2=\frac{1}{2} \dot{\phi}^2+V
-\frac{1}{2}\beta_A M^2 A_0^2\,,
\label{back1}\\
& &
-2M_{\rm pl}^2 \dot{H}
=\dot{\phi}^2
+\frac{1}{2}\beta_m M \dot{\phi}A_0\,,
\label{back2}\\
& &
\ddot{\phi}+3H \dot{\phi}+V_{,\phi}+
\frac{1}{2}M \beta_m \left( \dot{A}_0
+3H A_0 \right)=0\,,
\label{back3}\\
& &
A_0=-\frac{\beta_m}
{2 \beta_A M} \dot{\phi}\,,
\label{back4}
\ea
where $H \equiv \dot{a}/a$ is the Hubble expansion rate,  
a dot represents a derivative with respect to cosmic time $t$, and 
$V_{,\phi} \equiv dV/d\phi$.
{}From Eq.~(\ref{back4}), we notice that the temporal vector component 
$A_0$ is simply proportional to $\dot{\phi}$. 
Substituting Eq.~(\ref{back4}) into 
Eqs.~(\ref{back1}), (\ref{back2}), and (\ref{back3}),
we obtain 
\ba
& & 
3M_{\rm pl}^2 H^2=\frac{1}{2}\beta \dot{\phi}^2+V\,,
\label{back1d}\\
& &
-2M_{\rm pl}^2 \dot{H}=\beta \dot{\phi}^2\,,
\label{back2d}\\
& &
\ddot{\phi}+3H \dot{\phi}+\frac{V_{,\phi}}{\beta}=0\,,
\label{back3d}
\ea
where we have defined
\be
\beta \equiv 1-\frac{\beta_m^2}{4\beta_A}\,.
\label{beta}
\ee
The coupling $\beta$ is different from 1 due to the 
mixing term $\beta_m$. 
This leads to the modified evolution of $\phi$ 
compared to the standard case ($\beta=1$).

In Refs.~\cite{HKT18a,KT18}, the authors derived conditions for the absence of 
ghost and Laplacian instabilities of linear cosmological 
perturbations in the small-scale limit. 
The propagation speeds of tensor, vector, and scalar perturbations
are all equivalent to that of light for the theory given by the action 
(\ref{action}). The no-ghost conditions of tensor and vector perturbations 
are trivially satisfied, while the scalar ghost is absent under 
the condition
\be
q_s=\frac{M^2}{16} \left( 4\beta_A-\beta_m^2 \right)>0\,,
\ee
and hence $4\beta_A>\beta_m^2 \geq 0$. 
Then, the coupling (\ref{beta}) lies in the range 
\be
0 <\beta \le 1\,.
\label{betacon}
\ee
{}From Eq.~(\ref{back3d}),  the nonvanishing mixing term 
$\beta_m$ effectively leads to a faster inflaton velocity. 

Employing the slow-roll approximations
$\beta \dot{\phi}^2/2 \ll V$ and 
$|\ddot{\phi}| \ll |3H \dot{\phi}|$ in 
Eqs.~(\ref{back1d}) and (\ref{back3d}), 
it follows that 
\ba
& &
3M_{\rm pl}^2H^2 \simeq V\,,\label{sap1}\\
& &
3H \dot{\phi} \simeq -\frac{V_{,\phi}}{\beta}\,.
\label{sap2}
\ea
The slow-roll parameter associated with the 
cosmic expansion rate is given by  
\be
\epsilon \equiv -\frac{\dot{H}}{H^2} 
\simeq \frac{\epsilon_V}{\beta}\,,
\label{epsilon}
\ee
where we used Eq.~(\ref{back2d}), and defined 
\be
\epsilon_V \equiv \frac{M_{\rm pl}^2}{2} \left( 
\frac{V_{,\phi}}{V} \right)^2\,.
\ee
The existence of the nonvanishing mixing term $\beta_m$ breaks 
the relation $\epsilon \simeq \epsilon_V$ in standard inflation. 
The field value $\phi=\phi_f$ at the end of inflation can be 
derived by the condition $\epsilon(\phi_f)=1$, \emph{i.e.}, 
\be
\epsilon_V (\phi_f)=\beta\,.
\label{epVf}
\ee
The number of \emph{e}-foldings counted 
to the end of inflation is given by 
\be
N=\int_{\phi}^{\phi_f} \frac{H}{\dot{\tilde{\phi}}} d\tilde{\phi}
\simeq \frac{\beta}{M_{\rm pl}^2} \int_{\phi_f}^{\phi}
\frac{V}{V_{,\tilde{\phi}}} d\tilde{\phi}\,,
\label{efold}
\ee
where, in the last approximate equality, we again used the slow-roll 
approximation.
For smaller $\beta$, the number of \emph{e}-foldings gets smaller 
with a given initial value of $\phi$.
This is attributed to the fact that the inflaton velocity
 is effectively increased by the nonvanishing 
 coupling $\beta_m$.

If we introduce a rescaled field $\varphi$ defined by 
\be
\frac{d\varphi}{d\phi}=\sqrt{\beta}\,,
\label{dphi}
\ee
then Eqs.~(\ref{back1d}), (\ref{back2d}), (\ref{back3d}) 
reduce, respectively, to 
\ba
& &
3M_{\rm pl}^2 H^2=\frac{1}{2} \dot{\varphi}^2+V\,,\\
& &
-2M_{\rm pl}^2 \dot{H}= \dot{\varphi}^2\,,\\
& &
\ddot{\varphi}+3H \dot{\varphi}+V_{,\varphi}=0\,.
\ea
This means that the background dynamics in the presence of 
$\phi$ and $A_0 \propto \dot{\phi}$ is equivalent to the effective single-field 
dynamics driven by the scalar field $\varphi$. 
{}From Eq.~(\ref{dphi}), we have 
$\dot{\phi}=\dot{\varphi}/\sqrt{\beta}$, so the inflaton
$\phi$ evolves faster than the rescaled field $\varphi$ 
for $\beta_m \neq 0$.

\section{Tensor and vector perturbations}
\label{tvsec}

In this section, we revisit the tensor power spectrum generated  
during inflation \cite{HKT18a,KT18} and also discuss the evolution 
of vector perturbations in SVT theories given by the action 
(\ref{action}).

\subsection{Tensor perturbations}

The perturbed line element containing intrinsic tensor modes 
$h_{ij}(t, x^i)$ on the flat FLRW background is given by 
\be
ds_t^{2}=-dt^{2}+a^{2}(t) \left( \delta_{ij}
+h_{ij} \right)dx^i dx^j\,,
\ee
where $h_{ij}$ obeys the transverse and traceless conditions 
$\nabla^j h_{ij}=0$ and ${h_i}^i=0$. 
{}From Eq.~(3.2) of Ref.~\cite{HKT18a}, the second-order 
action of $h_{ij}$, for the theory given by Eq.~(\ref{action}), 
is the same as that in GR, \emph{i.e.}, 
\be
{\cal S}_t^{(2)}=\int dt d^3 x\,
\frac{a^3 M_{\rm pl}^2}{8} \delta^{ik} \delta^{jl} 
\left[ \dot{h}_{ij} \dot{h}_{kl}
-\frac{1}{a^2} (\partial h_{ij})
(\partial h_{kl})  \right]\,,
\label{St2}
\ee
where the symbol $\partial$ represents the spatial 
partial derivative. 
In Fourier space with the coming wavenumber $k$, the equation of 
motion of $h_{ij}$ is given by 
\be
\ddot{h}_{ij}+3H \dot{h}_{ij}+\frac{k^2}{a^2} h_{ij}
=0\,. \label{hij}
\ee
Deep inside the Hubble radius ($k/a \gg H$), the tensor 
perturbation is in a Bunch-Davies vacuum state, whereas 
after the Hubble exit ($k/a< H$) during inflation, 
$h_{ij}$ soon approaches a constant. 
Taking into account two polarization states, 
the primordial tensor power spectrum 
(per unit logarithmic wavenumber interval) 
generated during inflation yields \cite{HKT18a}
\be
{\cal P}_t=\frac{2H^2}{\pi^2 M_{\rm pl}^2}
\biggr|_{k=aH}\,,
\label{Pt0}
\ee
which should be evaluated at the Hubble exit. 
By using the slow-roll approximation (\ref{sap1}), 
Eq.~(\ref{Pt0}) can be expressed in terms of $V$, as 
\be
{\cal P}_t \simeq \frac{2V}{3\pi^2 M_{\rm pl}^4}
\biggr|_{k=aH}\,.
\label{Pt}
\ee
\subsection{Vector perturbations}

For the vector sector, we choose the perturbed line 
element in the flat gauge
\be
ds_v^2=-dt^2+2V_i dt dx^i+a^2(t) \delta_{ij} dx^i dx^j\,,
\ee
where the vector perturbation $V_i (t, x^i)$ obeys the transverse 
condition $\nabla^i V_i=0$. 
The spatial component of $A_{\mu}$ contains 
the intrinsic vector mode $Z_i$ and
the longitudinal scalar perturbation $\psi$, such that 
\be
A_i=Z_i+\nabla_i \psi\,,
\label{Ai}
\ee
where $Z_i$ obeys the condition $\nabla^i Z_i=0$.
In this section, we study the evolution of vector 
perturbations $Z_i$ during inflation, leaving the analysis 
of scalar mode $\psi$ for Sec.~\ref{persec}. 

Without loss of generality, we can choose the components 
of $V_i$ and $Z_i$ in the forms $V_i=(V_1(t,z), V_2(t,z),0)$ and $Z_i=(Z_1(t,z), Z_2(t,z),0)$.
After integrating out the nondynamical field $V_i$, the 
second-order action of vector perturbations 
reduces to \cite{HKT18a}
\be
{\cal S}_v^{(2)}=\int dt d^3 x \sum_{i=1}^2 
\frac{a}{2} \left[ \dot{Z}_i^2-\frac{1}{a^2} 
\left( \partial Z_i \right)^2-\beta_A M^2 Z_i^2 
\right]\,.
\label{Sv2}
\ee
Then, in Fourier space, the dynamical perturbation $Z_i$ obeys 
\be
\ddot{Z}_i+H \dot{Z}_i+\left( \frac{k^2}{a^2}
+\beta_A M^2 \right) Z_i=0\,,
\label{Zieq}
\ee
which can be written as 
\be
Z_i''+\left( k^2+a^2 \beta_A M^2 \right)Z_i=0\,,
\ee
where a prime represents the derivative with respect to 
the conformal time $\tau=\int a^{-1}dt$.
For the modes satisfying the condition 
$k^2 \gg a^2 \beta_A M^2$, the perturbation is 
in a Bunch-Davies vacuum state characterized 
by $Z_i=e^{-ik \tau}/\sqrt{2k}$. 
On the other hand, after the mass term $a^2 \beta_A M^2$ 
dominates over $k^2$ during inflation, we solve Eq.~(\ref{Zieq}) for
$Z_i$ under the conditions that $H={\rm constant}$ 
and that $k^2/a^2$ is negligible relative 
to $\beta_A M^2$. 
We then obtain the following solution 
\be
Z_i=A_{+} e^{\lambda_{+} t}+A_{-} e^{\lambda_{-} t}\,,
\ee
where $A_{\pm}$ are integration constants, and 
\be
\lambda_{\pm}=\frac{H}{2} 
\left[ -1 \pm \sqrt{1-\frac{4\beta_A M^2}{H^2}} \right]\,.
\label{lam}
\ee
Since $\beta_A>0$, the vector mass term leads to 
the exponential suppression of $Z_i$ after 
the perturbation enters the region $k^2/a^2<\beta_A M^2$. 
The term in the square root of Eq.~(\ref{lam}) 
becomes negative for $4\beta_A M^2>H^2$.
Now, we would like to consider the case in which $M$ is 
of the same order as the Hubble expansion rate $H$ 
during inflation.
Then, for the coupling  
\be
\beta_A={\cal O}(1)\,,
\label{beA}
\ee
the condition $4\beta_A M^2>H^2$ is satisfied. 
In this case, the amplitude of $Z_i$ 
decreases as 
\be
|Z_i| \propto e^{-Ht/2}\,,
\ee
with damped oscillations. 
Then, the vector perturbation decays very fast once 
it enters the region $k^2/a^2<\beta_A M^2$. 
Since $\beta_A M^2$ is of the same order as $H^2$, 
this exponential suppression starts to occur around 
the same moment of Hubble exit ($k^2/a^2<H^2$). 

In the following, we focus on the coupling $\beta_A$ 
of order 1. Then, the amplitude of vector 
perturbations at the end of inflation is completely negligible 
relative to those of tensor and scalar perturbations, so we can ignore 
the contributions of vector perturbations to the total 
primordial power spectrum.

\section{Primordial scalar power spectrum 
generated during inflation}
\label{persec}

Let us proceed to the derivation of the scalar power spectrum 
generated in our model given by the action (\ref{action}).
In doing so, we begin with the perturbed 
line-element on the FLRW background 
in the flat gauge:
\be
ds_s^{2}=-(1+2\alpha)\,dt^{2}+2\nabla_{i}\chi dt\,dx^{i}
+a^{2}(t) \delta_{ij}dx^i dx^j\,, 
\ee
where $\alpha$ and $\chi$ are scalar metric perturbations.
We decompose the scalar field $\phi$ into the background and 
perturbed parts, as 
\be
\phi=\phi_0(t)+\delta \phi(t,x^i)\,.
\ee
In the following, we omit the subscript ``0'' 
from the background value of $\phi$. 
The temporal component of $A^{\mu}$
is expressed in the form
\be
A^0=-A_0(t)+\delta A(t,x^i)\,,
\ee
whereas the spatial vector component $A_i$ contains the 
longitudinal scalar perturbation $\psi$ as Eq.~(\ref{Ai}).

The second-order action ${\cal S}_s^{(2)}$ of scalar perturbations 
was already computed in full parity-invariant SVT theories \cite{HKT18a}. 
In our theories given by the action (\ref{action}), 
we show the explicit form of ${\cal S}_s^{(2)}$ 
in Eq.~(\ref{Ssf}) of Appendix A.
Varying the action ${\cal S}_s^{(2)}$ with respect to 
$\alpha, \chi, \delta A$, we obtain the equations of motion 
for these nondynamical perturbations, see Eqs.~(\ref{ap1})-(\ref{ap3}). 
After integrating them out from the action, we are finally left
with two dynamical real fields, 
$\psi$ and $\delta \phi$. 
In general, any real scalar field ${\cal X}$ can be expanded in Fourier series, as 
\be
{\cal X}=\int \frac{d^3 k}{(2\pi)^3} \left[ {\cal X}_k (t) 
a({\bm k})e^{i {\bm k}\cdot {\bm x}}
+{\cal X}^*_k (t) a^\dagger({\bm k})e^{-i {\bm k} \cdot {\bm x}}\right]\,,
\ee
where ${\bm k}$ is a coming wavenumber and 
${\cal X}_k (t)$ is the mode function 
in Fourier space. For a quantized field ${\cal X}$, the coefficient
$a({\bm k})$ and its Hermitian conjugate $a^\dagger({\bm k})$ 
correspond to annihilation and creation operators.

Thus, the second-order action for dynamical 
perturbations ${\cal X}^{t}=(\psi_k, \delta \phi_k)$ in Fourier 
space can be written as
\be
{\cal S}_s^{(2)}=\int dt d^3x\,a^{3}\left( 
\dot{\vec{\mathcal{X}}}^{t}{\bm K}\dot{\vec{\mathcal{X}}}
-\frac{k^2}{a^2}\vec{\mathcal{X}}^{t}{\bm G}\vec{\mathcal{X}}
-\vec{\mathcal{X}}^{t}{\bm M}\vec{\mathcal{X}}
\right)\,,
\label{Ss2}
\ee
where   
${\bm K}$, ${\bm G}$, and ${\bm M}$
are $2 \times 2$ matrices. The matrix ${\bm M}$ does not 
contain the $k^2$ term. We note that the term 
$\vec{\mathcal{X}}^{t}{\bm B}\dot{\vec{\mathcal{X}}}$ 
appearing in Ref.~\cite{HKT18a} has been absorbed into 
${\bm M}$ after the integration by parts. 
The nonvanishing matrix components are given 
by~\footnote{Unlike Ref.~\cite{HKT18a}, the small-scale limit 
$k^2 \to \infty$ is not taken here, so that the components 
of ${\bm K}$ contain $k^2$-dependent terms.}
\ba
& &
K_{11}=\frac{k^2 \beta_A M^2}{2(k^2+a^2 \beta_A M^2)}\,,
\quad 
K_{12}=K_{21}=
\frac{\beta_m}{2\beta_A M}K_{11}\,,\nonumber \\
& &
K_{22}=\frac{1}{2}-\frac{a^2\beta_m^2 M^2}{8(k^2+a^2 \beta_A M^2)}
\,,\nonumber \\
& &
G_{11}=\frac{\beta_A M^2}{2}\,,\quad
G_{12}=G_{21}=\frac{\beta_m M}{4}\,,\quad
G_{22}=\frac{1}{2}\,,\nonumber \\
& &
M_{22}=\frac{V_{,\phi \phi}}{2}
-\frac{(1-\delta_{\phi}^2)V_{,\phi}^2}
{6M_{\rm pl}^2H^2}
-\frac{(1+\delta_{\phi})^4 V_{,\phi}^4}
{324 \beta H^6 M_{\rm pl}^4}\,,
\label{mass}
\ea
where we used the background Eqs.~(\ref{back1d})-(\ref{back3d}) 
to eliminate $\dot{H}$ and $\ddot{\phi}$. We also introduced 
the dimensionless quantity
\be
\delta_{\phi} \equiv \frac{\beta \ddot{\phi}}{V_{,\phi}}
=-\frac{3\beta H \dot{\phi}+V_{,\phi}}{V_{,\phi}}\,,
\ee
which is smaller than order 1 during inflation.
The off-diagonal components $K_{12}$ and $G_{12}$ 
do not vanish for $\beta_m \neq 0$.

To study the evolution of perturbations 
$\psi_k$ and $\delta \phi_k$ in Fourier space, 
we introduce the following combination 
\be
\delta \chi_k \equiv \psi_k+\frac{\beta_m}{2\beta_ A M} 
\delta \phi_k\,.
\label{delchi}
\ee
Varying the action (\ref{Ss2}) with respect to $\psi_k$ and 
using the properties that both $K_{12}/K_{11}$ and 
$G_{12}/G_{11}$ are equivalent to $\beta_m/(2\beta_ A M)$, 
we obtain 
\be
\frac{1}{a^3} \frac{d}{dt} 
\left( a^3 K_{11} \dot{\delta \chi}_k \right)
+\frac{k^2}{a^2} G_{11} \delta \chi_k=0\,.
\label{chieq}
\ee

For $k^2/a^2 \gg \beta_A M^2$, we have
$K_{11} \to \beta_A M^2/2=G_{11}$ and hence 
Eq.~(\ref{chieq}) reduces to 
\be
\ddot{\delta \chi}_k+3H \dot{\delta \chi}_k
+\frac{k^2}{a^2} \delta \chi_k=0\,.
\label{chieq1}
\ee
This equation is of the same form as Eq.~(\ref{hij}) 
for tensor perturbations, \emph{i.e.}, the equation of motion of 
a massless field.
For the modes deep inside the Hubble 
radius ($k^2/a^2 \gg H^2$), the canonically normalized field 
$v_k=\sqrt{2}a \delta \chi_k$ is
in a Bunch-Davies vacuum state characterized by 
$v_k=e^{-ik \int dt/a}/\sqrt{2k}$. 
Since we are considering the coupling in Eq.~(\ref{beA}) with 
$M \simeq H$ during inflation, the transition to 
another regime $k^2/a^2<\beta_A M^2$ occurs around 
the exit of Hubble radius.

For $k^2/a^2 \ll \beta_A M^2$, we have 
$K_{11} \to k^2/(2a^2)$, so Eq.~(\ref{chieq}) yields
\be
\ddot{\delta \chi}_k+H \dot{\delta \chi}_k
+\beta_A M^2 \delta \chi_k=0\,,
\label{chieq2}
\ee
which is of the same form as Eq.~(\ref{Zieq}) after taking the same limit. 
On the quasi de Sitter background 
($H\simeq {\rm constant}$), the solution to 
Eq.~(\ref{chieq2}) is given by 
\be
\delta \chi_k=A_{+} e^{\lambda_{+} t}
+A_{-} e^{\lambda_{-} t}\,,
\label{dchiso}
\ee
where $\lambda_{\pm}$ are equivalent to those given 
in Eq.~(\ref{lam}). 
Analogous to the intrinsic vector mode $Z_i$, the 
perturbation $\delta \chi_k$ starts to be 
exponentially suppressed after it enters the region 
$k^2/a^2<\beta_A M^2$.

For the coupling $\beta_A$ satisfying 
$4\beta_A M^2 > H^2$, 
the amplitude of $\delta \chi_k$ decreases as 
$|\delta \chi_k| \propto e^{-Ht/2}$. 
Then, the perturbation $\delta \chi_k$ is vanishing small 
at the end of inflation, so we can set $\delta \chi_k \simeq 0$ 
in Eq.~(\ref{delchi}) and hence 
\be
\psi_k \simeq -\frac{\beta_m}{2\beta_A M} \delta \phi_k\,.
\label{psiphieq}
\ee
One can notice that, from Eq.~(\ref{back4}), the relation 
between $\psi_k$ and $\delta \phi_k$ is analogous to that 
between $A_0$ and $\dot{\phi}$.

The only possibility for avoiding the above 
strong suppression is to consider 
the small coupling $\beta_A \ll 1$. 
In this case, there is a period characterized by 
$H^2>k^2/a^2>\beta_A M^2$ during which the perturbation 
$\delta \chi_k$ is temporally frozen with the value at Hubble
radius crossing. However, after the perturbation enters 
the region $k^2/a^2<\beta_A M^2$, $\delta \chi_k$ starts 
to decay according to Eq.~(\ref{dchiso}). 
It is possible to derive the solution to Eq.~(\ref{chieq2}) 
even for the background where the scale factor evolves
as $a \propto t^p$, where $p$ is a positive 
constant. In this case the resulting solution is given by 
$|\delta \chi_k| \propto t^{-p/2}$, so the suppression of 
$\delta \chi_k$ also occurs after inflation whenever $H^2$ 
drops below the order of $\beta_A M^2$.

Varying the action (\ref{Ss2}) with respect to 
$\delta \phi_k$, it follows that 
\ba
& &
\frac{1}{a^3} \frac{d}{dt} \left[ a^3 \left( K_{22} 
\dot{\delta \phi}_k+K_{12} \dot{\psi}_k \right) \right]
+\frac{k^2}{a^2} \left( G_{22} \delta \phi_k+G_{12} 
\psi_k \right) \nonumber \\
& &
+M_{22} \delta \phi_k=0\,.
\label{delphieq}
\ea
Now, we employ Eq.~(\ref{delchi}) and its time derivative 
to eliminate $\psi_k$ and $\dot{\psi}_k$ from Eq.~(\ref{delphieq}).
In doing so, we also resort to the fact that $\delta \chi_k$ obeys Eq.~(\ref{chieq}).
Then, the contributions arising from $\delta \chi_k$ to Eq.~(\ref{delphieq}) 
cancel out, so that 
\be
\frac{1}{a^3} \frac{d}{dt} \left( a^3 \tilde{K}_{22} 
\dot{\delta \phi}_k \right)+\left( \frac{k^2}{a^2} \tilde{G}_{22}
+M_{22} \right) \delta \phi_k=0\,,
\label{delphieq2}
\ee
where 
\ba 
& &
\tilde{K}_{22} \equiv K_{22}-\frac{\beta_m}{2\beta_A M}K_{12}
=\frac{\beta}{2}\,,\label{K22} \\
& &
\tilde{G}_{22} \equiv G_{22}-\frac{\beta_m}{2\beta_A M}G_{12}
=\frac{\beta}{2}\,.
\label{KG}
\ea
Taking the limit $\beta \to 1$ in Eq.~(\ref{delphieq2}) 
with Eqs.~(\ref{K22}) and (\ref{KG}), we recover the 
perturbation equation of $\delta \phi_k$ in standard 
single-field inflation.

We introduce the canonically normalized field 
$\delta \sigma_k$, as 
\be
\delta \sigma_k \equiv a\sqrt{\beta} \delta \phi_k\,.
\ee
Then, we can express Eq.~(\ref{delphieq2}) in the form 
\be
\delta \sigma_k''+\left( k^2 -\frac{a''}{a}
+\frac{2a^2M_{22}}{\beta} \right) 
\delta \sigma_k=0\,.
\label{delphieq3}
\ee
On the quasi de-Sitter background characterized by 
$H \simeq {\rm constant}$, the conformal time 
$\tau=\int a^{-1} dt$ is 
approximately given by $\tau \simeq -(1+\epsilon)/(aH)$.
Applying  the slow-roll approximation (\ref{sap1}) to 
the mass term $M_{22}$ and picking up 
next-to-leading order terms in slow-roll in 
Eq.~(\ref{delphieq3}), we obtain
\be
\delta \sigma_k''+\left[ k^2 
-2(aH)^2 \left( 1+\frac{5\epsilon_V-3\eta_V}
{2\beta} \right) \right]\delta \sigma_k=0\,,
\label{delphieq4}
\ee
where we used the relation (\ref{epsilon}) and 
introduced the second slow-roll parameter 
\be
\eta_V \equiv \frac{M_{\rm pl}^2 V_{,\phi \phi}}{V}\,.
\ee
Neglecting the time variations of $\epsilon_V$ and $\eta_V$, 
the solution to Eq.~(\ref{delphieq4}), which recovers 
the Bunch-Davies vacuum state 
($\delta \sigma_k=e^{-i k\tau}/\sqrt{2k}$) 
in the asymptotic past ($k \tau \to -\infty$), is given by 
\be
\delta \sigma_k=\frac{\sqrt{\pi |\tau|}}{2}
e^{i(1+2\nu)\pi/4}H_{\nu}^{(1)} (k |\tau|)\,,
\ee
where $H_{\nu}^{(1)} (k |\tau|)$ is the Hankel function 
of first kind, and 
\be
\nu=\frac{3}{2}+\frac{3\epsilon_V-\eta_V}{\beta}\,.
\ee
Using the relations 
$H_{\nu}^{(1)} (k |\tau|) \to -(i/\pi)\Gamma(\nu)
(k|\tau|/2)^{-\nu}$ for $k \tau \to 0$ and 
$\Gamma (3/2)=\sqrt{\pi}/2$, the solution for  
$\delta \phi_k$ long after the Hubble exit 
during inflation is
\be
\delta \phi_k=i \frac{H (1-\epsilon)}{k^{3/2}\sqrt{2\beta}} 
\frac{\Gamma (\nu)}{\Gamma(3/2)}
\left( \frac{k |\tau|}{2} \right)^{3/2-\nu}\,.
\label{delphiso}
\ee
In the de-Sitter limit characterized by 
$\epsilon_V \to 0$ and $\eta_V \to 0$, the solution (\ref{delphiso})
reduces to $\delta \phi_k \to iH/(k^{3/2}\sqrt{2\beta})$.

We introduce the curvature perturbation in 
flat gauge incorporating both the field perturbations 
$\delta \phi_k$ and $\psi_k$, as \cite{HKT18b} 
\be
{\cal R}=-\frac{H (\dot{\phi} \delta \phi_k
+M^2 A_0 \psi_k)}
{\dot{\phi}^2+M^2 A_0^2}\,.
\label{curper}
\ee
By using Eq.~(\ref{back4}) and eliminating $\psi_k$ 
on account of Eq.~(\ref{delchi}), we can write 
Eq.~(\ref{curper}) in the form  
\be
{\cal R}={\cal R}_{\phi}+{\cal R}_{\chi}\,,
\ee
where 
\be
{\cal R}_{\phi}=
-\frac{H \delta \phi_k}{\dot{\phi}}\,,\qquad
{\cal R}_{\chi}=\frac{2\beta_m \beta_A}
{4\beta_A^2+\beta_m^2}
\frac{H M \delta \chi_k}{\dot{\phi}}\,.
\ee
Since $\delta \chi_k$ is exponentially suppressed 
by the end of inflation, we only need to compute the power 
spectrum of ${\cal R}_{\phi}$. 
Taking Eq.~(\ref{delphieq2}) with the mass term 
$M_{22}$ given in Eq.~(\ref{mass}), 
the perturbation ${\cal R}_{\phi}$ obeys
\be
\frac{1}{a^3  \epsilon} \frac{d}{dt} \left( a^3 \epsilon 
\dot{\cal R}_{\phi} \right)+\frac{k^2}{a^2}{\cal R}_{\phi}=0\,.
\ee
In the large-scale limit ($k^2/a^2 \to 0$), we obtain 
the following solution 
\be
{\cal R}_{\phi}=c_1+c_2 \int \frac{dt}{a^3 \epsilon}\,,
\label{Rphi}
\ee
where $c_1$ and $c_2$ are integration constants. 
In slow-roll inflation, the second term on 
the right hand side of Eq.~(\ref{Rphi}) can be identified as 
a decaying mode. Then, ${\cal R}_{\phi}$ 
approaches the constant $c_1$ soon after the Hubble exit. 
Then, the primordial power spectrum of ${\cal P}_{{\cal R}_{\phi}}$ 
per unit logarithmic wavenumber interval can be 
computed at $k=aH$, as 
\be
{\cal P}_{{\cal R}_{\phi}} \equiv \frac{k^3}{2\pi^2} 
\left| {\cal R}_{\phi} \right|^2
=\frac{H^4}{4\pi^2 \dot{\phi^2} \beta} \biggr|_{k=aH}\,,
\label{PRphi}
\ee
where we used the leading-order solution of Eq.~(\ref{delphiso}). 
Applying the slow-roll approximations (\ref{sap1})-(\ref{sap2})
to Eq.~(\ref{PRphi}) and neglecting the contribution from 
$\delta \chi_k$ to the total curvature perturbation ${\cal R}$, 
the resulting primordial scalar  power spectrum is given by 
\be
{\cal P}_{\cal R} \simeq \frac{\beta V^3}
{12\pi^2 M_{\rm pl}^6 V_{,\phi}^2} \biggr|_{k=aH}\,.
\label{PR}
\ee
In comparison with the canonical picture of single-field inflation,
the coupling $\beta$ induces different behavior 
for the scalar power spectrum.
Using the background field $\varphi$ defined by 
Eq.~(\ref{dphi}), the power spectrum (\ref{PR}) can 
be written in the form 
${\cal P}_{\cal R}=V^3/
(12\pi^2  M_{\rm pl}^6 V_{,\varphi}^2)|_{k=aH}$. 
This means that, as long as the perturbation $\delta \chi_k$ 
is negligibly small compared to $\delta \phi_k$ at the end of 
inflation, the effective single-field description in terms of 
$\varphi$ also works for curvature perturbations.

\section{Observational signatures in CMB}
\label{obsersec}

In this section, we compute inflationary observables
to confront our SVT theories with the CMB data of 
temperature anisotropies and study how they are 
modified by the presence of the coupling $\beta$.

\subsection{Inflationary observables}

In Sec.~\ref{tvsec}, we showed that vector perturbations 
are exponentially suppressed relative to scalar and tensor 
perturbations at the end of inflation, so we neglect the 
contribution of vector modes to the inflationary power spectra.
At the pivot wavenumber $k_0=0.05$ Mpc$^{-1}$, 
the amplitude of curvature perturbations constrained from 
\emph{Planck} 2018 observations is \cite{Planck2018}
\be
{\cal P}_{\cal R}=\frac{\beta V^3}
{12\pi^2 M_{\rm pl}^6 V_{,\phi}^2}=
2.1 \times 10^{-9}\,.
\label{PRk0}
\ee
The spectral indices of tensor and scalar perturbations 
are defined, respectively, by 
\ba
n_t &\equiv& \frac{d \ln {\cal P}_{t}}{d \ln k} 
\biggr|_{k=aH}\,,\\
n_s &\equiv& 1+\frac{d \ln {\cal P}_{\cal R}}{d \ln k} 
\biggr|_{k=aH}\,.
\ea
{}From Eqs.~(\ref{Pt}) and (\ref{PR}), we obtain 
\ba
n_t &=& -\frac{2\epsilon_V}{\beta}\,,
\label{nt}\\
n_s &=& 1-\frac{1}{\beta} \left( 6 \epsilon_V 
-2\eta_V \right)\,,
\label{ns}
\ea
where we used the slow-roll approximations 
(\ref{sap1})-(\ref{sap2}).
The tensor-to-scalar ratio is given by 
\be
r \equiv \frac{{\cal P}_t}{{\cal P}_{\cal R}}
=\frac{16\epsilon_V}{\beta}=16\epsilon\,.
\label{ratio}
\ee
From Eqs.~(\ref{nt}) and (\ref{ratio}), the following 
consistency relation holds
\be
r=-8n_t\,,
\ee
which is of the same form as that in standard single-field 
inflation. We study how the coupling $\beta$ 
modifies the observational prediction of $n_s$ and $r$. 
We show that this modification generally depends on the 
form of inflaton potentials.

\subsection{Different inflaton potentials and \emph{Planck} 2018 constraints}

In the following, we consider three different inflaton potentials 
arising in (i) natural inflation, (ii) $\alpha$-attractors, 
and (iii) brane inflation.
We also discuss whether these models can be consistent 
with the latest \emph{Planck} 2018 data \cite{Planck2018} in presence of 
the scalar-vector mixing.

\subsubsection{Natural inflation}

In natural inflation \cite{natural}, the potential is given by  
\be
V(\phi)=M^2 M_{\rm pl}^2 \left[ 1+\cos \left( \frac{\phi}{f} 
\right) \right]\,,
\ee
where $f$ is a mass scale associated with the shift symmetry.
In this case, the observables (\ref{PRk0}), (\ref{ns}), and 
(\ref{ratio}) reduce, respectively, to 
\ba
{\cal P}_{\cal R} &=&
\frac{f_{\beta}^2 M^2 (1+x)^2}
{12 \pi^2 M_{\rm pl}^2 (1-x)}=2.1 \times 10^{-9}\,,
\label{PRna} \\
n_s &=& 1-\frac{3-x}{f_{\beta}^2 (1+x)}\,,
\label{nsna}\\
r &=& \frac{8(1-x)}{f_{\beta}^2 (1+x)}\,,
\label{rna}
\ea
where $f_{\beta} \equiv \sqrt{\beta}f/M_{\rm pl}$ and 
$x \equiv \cos(\phi/f)$.
{}From Eq.~(\ref{efold}), we obtain 
$N=f_{\beta}^2 \ln [(1-x_f)/(1-x)]$, so that  
\be
x=1-(1-x_f)e^{-N/f_{\beta}^2}\,,
\label{xex}
\ee
where $x_f=(1-2f_{\beta}^2)/(1+2f_{\beta}^2)$ is the value 
of $x$ at the end of inflation determined by the condition 
(\ref{epVf}). Substituting Eq.~(\ref{xex}) into 
Eqs.~(\ref{nsna}) and (\ref{rna}), it follows that 
$n_s$ and $r$ depend on $f_{\beta}$ and $N$.
For a given $N$, these observables are functions of 
$f_{\beta}$ alone. Hence the theoretical curve in the 
$(n_s, r)$ plane is the same as that in standard 
natural inflation. The only difference is that 
the coupling $f/M_{\rm pl}$ is now modified to 
$f_{\beta}=\sqrt{\beta}f/M_{\rm pl}$. 
{}From \emph{Planck} 2015 data  \cite{Planck2015}, the coupling 
is constrained to be ${\rm log}_{10} (f_{\beta})>0.84$ 
at 95 \% CL, \emph{i.e.}, 
\be
f>\frac{6.9M_{\rm pl}}{\sqrt{\beta}}\,.
\ee
As in the standard case, the trans-Planckian problem about the scale $f$ 
also persists for $\beta<1$.
With given values of $f, \beta$, and $N$, the mass scale $M$ 
is known from the \emph{Planck} normalization (\ref{PRna}).

The recent \emph{Planck} 2018 data combined with the 
data of B-mode polarizations available from the 
BICEP2/Keck field (BK14) and baryon acoustic 
oscillations (BAO) indicate that most of the theoretical values 
of $n_s$ and $r$ in natural inflation are outside of the 
95 \% CL observational contour, see Fig.~8 of 
Ref.~\cite{Planck2018}. As shown above, 
this situation is not improved by the mixing term  
$\beta_m$ between inflaton and vector fields.

\subsubsection{$\alpha$-attractors}

The $\alpha$-attractor model \cite{alpha} is given by the potential 
\be
V(\phi)=\frac{3}{4}\alpha_c M^2 M_{\rm pl}^2
\left[ 1-\exp \left(-\sqrt{\frac{2}{3\alpha_c}} \frac{\phi}
{M_{\rm pl}} \right) \right]^2\,,
\label{alpo}
\ee
where $\alpha_c$ is a dimensionless 
constant~\footnote{We note that the same potential can be 
derived from Brans-Dicke theory with the Lagrangian 
$\mathcal{L}=M_{\rm pl}\phi R/2-V_0(\phi-M_{\rm pl})^2$ 
after a conformal transformation 
to the Einstein frame -- see Eq.~(109) of Ref.~\cite{Defelice11}. 
The observational constraints on this model were already 
performed in 2011 -- see Fig.~3 of Ref.~\cite{Defelice11}.}. 
Starobinsky inflation \cite{Sta80} characterized by the Lagrangian 
$f(R)=R+R^2/(6M^2)$ gives rise to the potential 
(\ref{alpo}) with $\alpha_c=1$ after a conformal transformation 
to the Einstein frame. In the limit that $\alpha_c \to \infty$, 
the potential (\ref{alpo}) reduces to that in chaotic 
inflation: $V(\phi)=M^2 \phi^2/2$.

\begin{figure}[h]
\begin{center}
\includegraphics[height=3.4in,width=3.4in]{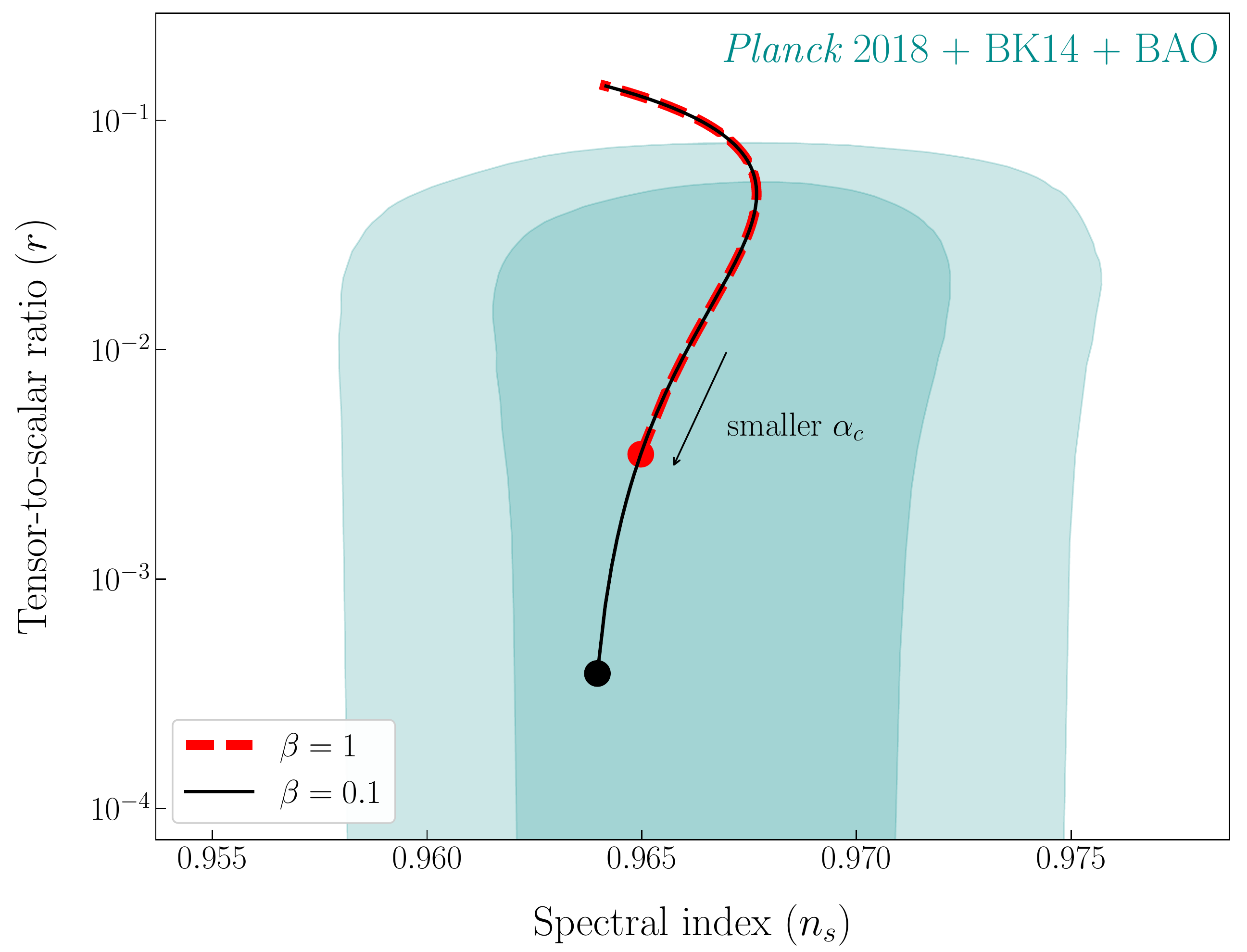}
\end{center}
\caption{\label{fig1}
Observational constraints on $\alpha$-attractors in the $(n_s, r)$ plane.
The green contours represent the 68 \% CL (inside) and 95 \% CL (outside) 
boundaries derived by the joint data analysis 
of \emph{Planck} 2018 $+$ BK14 $+$ BAO 
at $k=0.002$~Mpc$^{-1}$ \cite{Planck2018}. 
The red dashed and black thin solid lines correspond
to the cases $\beta=1$ and $\beta=0.1$, respectively, 
with $N=55$ and $1 \le \alpha_c \le 10^6$. 
The red and black circles represent Starobinsky inflation ($\alpha_c=1$) 
with $\beta=1$ and $\beta=0.1$, respectively. 
}
\end{figure}

For $\alpha$ attractors, the inflationary observables are 
\ba
{\cal P}_{\cal R} &=&
\frac{3\alpha_c^2 \beta M^2(1-y)^4}
{128 \pi^2 M_{\rm pl}^2 y^2}=2.1 \times 10^{-9}\,,
\label{PRal} \\
n_s &=& 1-\frac{8y(1+y)}{3\alpha_c \beta (1-y)^2}\,,
\label{nsal}\\
r &=& \frac{64y^2}{3\alpha_c \beta (1-y)^2}\,,
\label{ral}
\ea
where $y \equiv e^{-\sqrt{2/(3\alpha_c)}\phi/M_{\rm pl}}$.
The number of \emph{e}-foldings is given by 
\be
N=\frac{3}{4}\alpha_c \beta \left( 
\frac{1}{y}-\frac{1}{y_f}+\ln \frac{y}{y_f}
 \right)\,,
\label{Nat}
\ee
where $y_f=(3\alpha_c \beta-2\sqrt{3\alpha_c \beta})
/(3\alpha_c \beta-4)$ is the value of $y$ at the 
end of inflation.

For $\alpha_c<{\cal O}(10)$, 
$y$ is smaller than order 1 during inflation.
In this case, the dominant contribution to $N$ is the first term 
in the parenthesis of Eq.~(\ref{Nat}), \emph{i.e.}, 
$y \simeq 3\alpha_c \beta/(4N) \ll 1$. 
Substituting this expression into Eqs.~(\ref{nsal}) 
and (\ref{ral}), we obtain 
\be
n_s \simeq 1-\frac{2}{N}\,,\qquad 
r \simeq \frac{12\alpha_c \beta}{N^2}\,.
\label{atob1}
\ee
While $n_s$ does not depend on $\beta$, the scalar-vector 
mixing ($\beta_m \neq 0$) leads to a smaller value for the tensor-to-scalar 
ratio compared to the case $\beta=1$. 
The \emph{Planck} normalization (\ref{PRal}) gives 
\be
M=1.3 \times 10^{-5}M_{\rm pl}\sqrt{\beta} 
\left( \frac{55}{N} \right)\,,
\ee
so that $M$ decreases for smaller $\beta$.

For $\alpha_c \gg {\cal O}(10)$, $y$ approaches 1 with 
increasing $\alpha_c$.
Expansion of Eq.~(\ref{Nat}) around $y=1$ shows that 
the number of \emph{e}-foldings long before the end of inflation is
approximately given by $N \simeq 3\alpha_c \beta (1-y)^2/8 \gg 1$.
In this regime, the observables (\ref{nsal}) and (\ref{ral}) reduce to
\be
n_s \simeq 1-\frac{2}{N}\,,\qquad 
r \simeq \frac{8}{N}\,,
\label{atob2}
\ee
which are equivalent to those in standard chaotic inflation 
driven by the potential $V(\phi)=M^2 \phi^2/2$ \cite{TOKA}. 
{}From Eq.~(\ref{atob2}), the coupling $\beta$ modifies 
neither $n_s$ nor $r$ for $\alpha_c \gg {\cal O}(10)$.

In Fig.~\ref{fig1}, we plot the theoretical curves in the $(n_s,r)$ plane 
for $\beta=1$ (red dashed) and $\beta=0.1$ (black thin solid) for $N=55$ and 
$1 \le \alpha_c \le 10^6$. For $\alpha_c \gg {\cal O}(10)$, the observables 
converge to the values (\ref{atob2}) irrespective of the coupling $\beta$.
With decreasing $\alpha_c$, the difference of $r$ 
between the two different values of $\beta$ tends to be significant. 
In Starobinsky inflation ($\alpha_c=1$), for example, we have 
$r=3.9 \times 10^{-4}$ for 
$\beta=0.1$. As estimated from Eq.~(\ref{atob1}), this is by 
one order of magnitude smaller than the value $r=3.5 \times 10^{-3}$ 
for $\beta=1$. In both cases, the models are inside 68 \% CL observational 
contour constrained from \emph{Planck} 2018 $+$ BK14 $+$ BAO data. 
Interestingly, even if future observations place the upper limit 
of $r$ down to $10^{-3}$, the model with $\alpha_c=1$ can be still rescued 
by the coupling $\beta$.

As we observe in Fig.~\ref{fig1}, the scalar spectral index $n_s$ for 
$\beta=0.1$ and $\alpha_c=1$ is slightly smaller than that for 
$\beta=1$ and $\alpha_c=1$. 
This reflects the fact that, in the latter case, the approximation $y \ll 1$ 
we used for the derivation of $n_s$ in Eq.~(\ref{atob1}) is not 
completely accurate. As the product $\alpha_c \beta$ decreases 
toward 0, the observables approach $n_s \to 1-2/N$ and $r \to 0$, 
which are favored in current CMB observations.

Since the coupling $\beta$ smaller than 1 can reduce the value of $r$, 
the bound on $\alpha_c$ is less stringent compared to the case $\beta=1$. 
For $\beta=1$ the observational upper limit is  
$\alpha_c<4.4 \times 10$ (68 \% CL), while, for $\beta=0.1$, 
the bound is loosened: $\alpha_c<4.2 \times 10^2$ (68 \% CL). 
Unless $\alpha_c$ is very much larger than 1 to approach 
the asymptotic values of $n_s$ and $r$ given by Eq.~(\ref{atob2}), 
the product $\alpha_c \beta$ is constrained to be 
\be
\alpha_c \beta \lesssim 40\,, 
\ee
at 68 \% CL.
The main reason why $r$ is reduced by the mixing term $\beta_m$ 
is that the coupling $\beta$ leads to smaller 
$y \simeq 3\alpha_c \beta/(4N)$ (\emph{i.e.}, larger $\phi$) 
for $\alpha_c<{\cal O}(10)$. This effect overwhelms 
the coupling $\beta$ in the denominator of Eq.~(\ref{ral}), 
so that $r$ has the dependence $r \propto \alpha_c \beta/N^2$. 
In other words, for $\beta<1$, we require that inflation occurs 
in the region where the potential is flatter relative to the case 
$\beta=1$ to acquire the same number of \emph{e}-foldings. 
This effectively reduces the value of $r=16 \epsilon$ 
for given $N$.

\subsubsection{Brane inflation}

Finally, we study brane inflation characterized by 
the effective potential 
\be
V(\phi)=M^2 M_{\rm pl}^2 \left[ 1-\left( 
\frac{\mu}{\phi} \right)^p+\cdots \right]\,,
\label{Bpo}
\ee
where $p$ and $\mu$ are positive constants. 
The models arising from the setup of D-brane and anti D-brane 
configuration have the power $p=2$ \cite{branep=2} 
or $p=4$ \cite{branep=4a,branep=4b}. 
For the positivity of $V(\phi)$, we require that 
$z \equiv \phi/\mu>1$.  
We assume that inflation ends around 
$\phi \approx \mu$ before 
the additional terms denoted by the ellipsis in Eq.~(\ref{Bpo}) 
contributes to the potential.

The observables (\ref{PRk0}), (\ref{ns}), and 
(\ref{ratio}) reduce, respectively, to 
\ba
{\cal P}_{\cal R} &=&
\frac{\beta M^2 \mu^2 (z^p-1)^3}
{12 \pi^2 M_{\rm pl}^4p^2 z^{p-2}}=2.1 \times 10^{-9}\,,
\label{PRD} \\
n_s &=& 1-\frac{pM_{\rm pl}^2[2(p+1)z^p+p-2]}
{\mu^2 z^2 (z^p-1)^2 \beta}\,,
\label{nsD}\\
r &=& \frac{8p^2 M_{\rm pl}^2}
{\mu^2 z^2 (z^p-1)^2 \beta}\,.
\label{rD}
\ea
The number of \emph{e}-foldings is given by 
\be
N \simeq \frac{\beta \mu^2[z^2 (2z^p-p-2)+p]}
{2M_{\rm pl}^2 p(p+2)}\,,
\label{ND}
\ee
where we used the fact that the value of $z$ at the end 
of inflation is $z_f \simeq 1$.

Since inflation occurs in the region $z^p \gg 1$, we pick up 
the dominant contributions to Eqs.~(\ref{nsD}), (\ref{rD}), 
and (\ref{ND}). Then we have 
$z^{p+2} \simeq M_{\rm pl}^2 p(p+2)N/\beta \mu^2$, and 
\ba
n_s &\simeq& 1-\frac{2(p+1)}{(p+2)N}\,,\label{nsDb}\\
r &\simeq& 8p^2 \left( \frac{\beta \mu^2}{M_{\rm pl}^2} 
\right)^{\frac{p}{p+2}} 
\left[ \frac{1}{p(p+2)N} \right]^{\frac{2(p+1)}{p+2}}\,,
\label{rDb}
\ea
which show that the $\beta$ dependence appears in 
$r$ but not in $n_s$.
{}From Eq.~(\ref{nsDb}), we obtain $n_s=1-3/(2N)$ for $p=2$ and 
$n_s=1-5/(3N)$ for $p=4$, so they are larger than 
$n_s$ in Eq.~(\ref{atob1}) of $\alpha$ attractors.
{}From Eq.~(\ref{rDb}), the tensor-to-scalar ratio has 
the dependence 
$r \propto \beta^{1/2}/N^{3/2}$ for $p=2$ and 
$r \propto \beta^{2/3}/N^{5/3}$ for $p=4$.
In the limit that $p \gg 1$, we have 
$n_s \simeq 1-2/N$ and $r \propto \beta/N^2$, so 
they have the same dependence of $N$ and 
$\beta$ as those in $\alpha$-attractors with 
$\alpha_c<{\cal O}(10)$. 
The scalar-vector mixing works to 
reduce the tensor-to-scalar ratio compared to 
the case $\beta=1$. 
Unlike $\alpha$-attractors in which the dependence of 
$r$ with respect to $\beta$ depends on $\alpha_c$, 
the reduction of $r$ induced by the coupling $\beta$ occurs irrespective of 
the values of $\mu$.

\begin{figure}[h]
\begin{center}
\includegraphics[height=3.4in,width=3.4in]{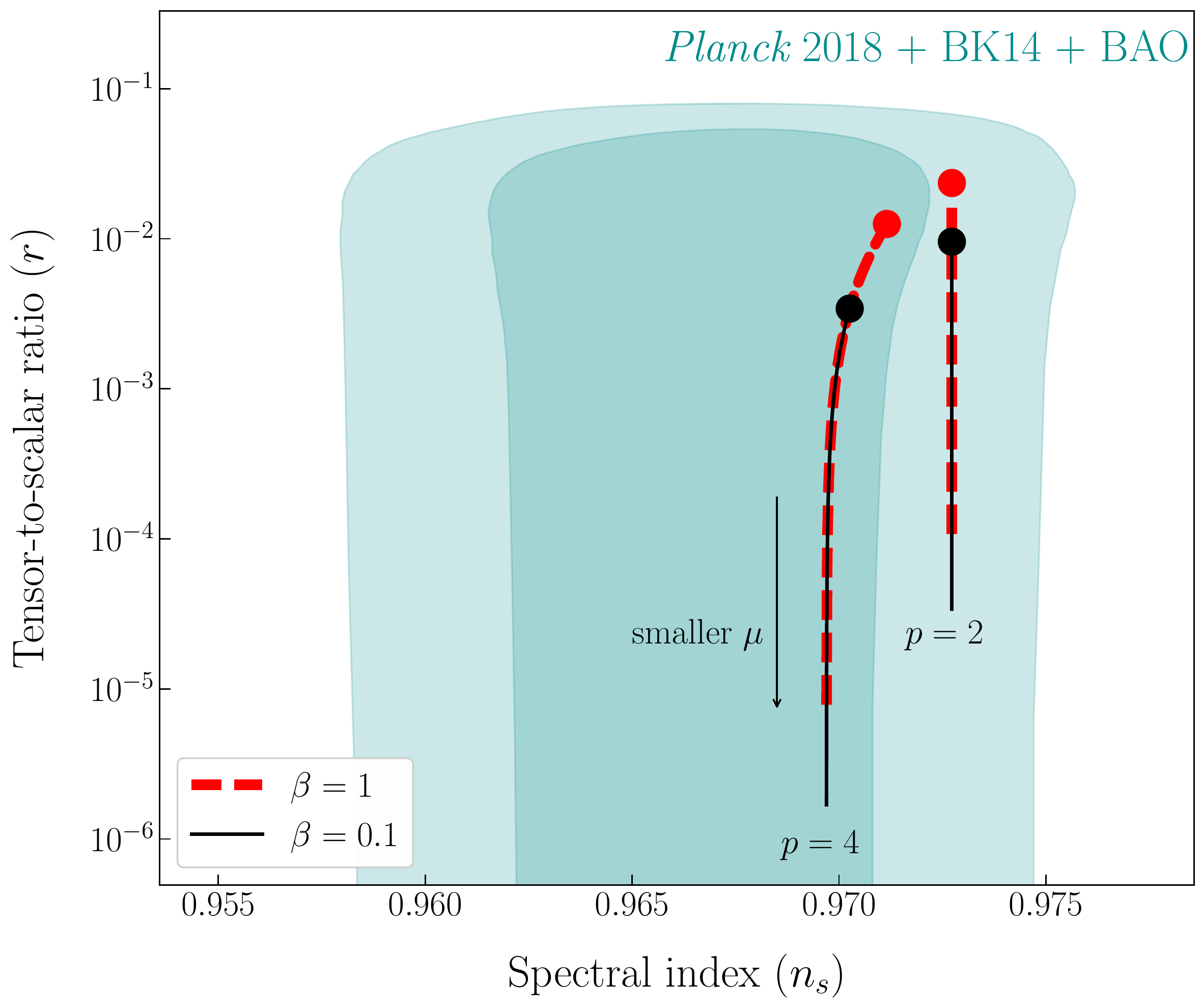}
\end{center}
\caption{\label{fig2}
Observational constraints on brane inflation in the 
$(n_s, r)$ plane for $p=2$ and $p=4$.
The green contours are the same as those in Fig.~\ref{fig1}.
The red dashed and black thin solid lines represent the cases 
$\beta=1$ and $\beta=0.1$, respectively, 
with $N=55$ and $-1.5 \le {\rm log}_{10}(\mu/M_{\rm pl}) \le 1.0$. 
The red and black circles correspond to 
${\rm log}_{10}(\mu/M_{\rm pl})=1.0$
with $\beta=1$ and $\beta=0.1$, respectively. 
}
\end{figure}

In Fig.~\ref{fig2}, we plot the theoretical curves in the 
$(n_s, r)$ plane for the brane inflation scenario with $\beta=1$ and 
$\beta=0.1$ for the mass range between 
$10^{-3/2} \le \mu/M_{\rm pl} \le 10$.  
We consider the models with two different powers: 
$p=2$ and $p=4$. 
For smaller $\mu$, $z$ gets larger and hence 
the approximate results (\ref{nsDb})-(\ref{rDb}) 
tend to be more accurate. 
As estimated from Eq.~(\ref{nsDb}), the scalar spectral 
index is nearly constant, \emph{i.e.}, $n_s \simeq 0.9727$ for $p=2$ 
and $n_s \simeq 0.9697$ for $p=4$. 

The red circle plotted on the line for $p=2$ of Fig.~\ref{fig2} 
corresponds to the model parameters 
$\beta=1$ and $\mu/M_{\rm pl}=10$, in which case 
the model is inside the 95~\% CL observational contour 
with $r=2.35 \times 10^{-2}$. 
{}From Eq.~(\ref{rDb}), the tensor-to-scalar ratio 
decreases for smaller values of $\beta$ and $\mu$. 
When $p=2$, $\beta=0.1$, $\mu/M_{\rm pl}=10$, 
the numerical value of $r$ is given by 
$9.53 \times 10^{-3}$ ---see the black circle on the line for $p=2$ 
of Fig.~\ref{fig2}. The models with $\beta<1$ and 
$\mu \lesssim 10M_{\rm pl}$ are consistent with 
the current upper bound of $r$. For $p=2$, 
the scalar spectral index is between the 68~\% CL and 
95~\% CL observational boundaries.

The model with $p=4$ gives rise to $n_s$ smaller than 
that for $p=2$, so the former model enters the 
68\% CL observational contour for 
$\mu \lesssim 10M_{\rm pl}$ and $\beta \leq 1$.
The red circle shown on the line for $p=4$ of Fig.~\ref{fig2} 
corresponds to $\beta=1$ and $\mu/M_{\rm pl}=10$, 
in which case $r=1.25 \times 10^{-2}$. 
For $\beta=0.1$, this value is reduced to 
$r=3.41 \times 10^{-3}$. 
For smaller $\beta$ and $\mu$, the tensor-to-scalar 
ratio approximately decreases as 
$r \propto (\beta \mu^2)^{2/3}$ for $p=4$. 

We note that the increase of $r$ induced by the coupling 
$\beta~(<1)$ in the denominator of Eq.~(\ref{rD}) is 
switched to the decrease of $r$ by the other term 
$z^{2+2p} \propto \beta^{-(2+2p)/(2+p)}$. 
Analogous to $\alpha$-attractors with 
$\alpha_c<{\cal O}(10)$, this behavior occurs in small-field inflation 
in which the variation of $\phi$ during inflation does not exceed the order of 
$M_{\rm pl}$. 
In $\alpha$-attractors with $\alpha_c \gg {\cal O}(10)$, which 
corresponds to large-field inflation, the decrease of $r$ 
induced by $\beta$ is not significant.
In chaotic inflation (the limit $\alpha_c \to \infty$ in $\alpha$-attractors),  
both $\epsilon_V$ and $\eta_V$ are inversely proportional to $N$, 
in which case both $n_s$ and $r$ solely depend on $N$ 
but not on $\beta$. 
In small-field inflation, 
$\epsilon_V$ and $\eta_V$ have different $N$ dependence 
with $\epsilon_V \ll |\eta_V|$, in which case the explicit 
$\beta$ dependence appears in $r$.

\section{Conclusions}
\label{consec}

This work was devoted to the study of prominent effective field theories with helicity-0 
and helicity-1 fields in the presence of a dimension-3 operator that couples the two sectors. 
We have investigated the implications of this coupling for inflation driven by the helicity-0 
mode with a given potential energy, paying particular attention to the evolution of 
cosmological perturbations. 
At the background level, the temporal component of helicity-1 mode, $A_0$, is just 
an auxiliary (nondynamical) field, so that it can be directly integrated out 
in terms of the time derivative of helicity-0 mode. 
In this way, the background dynamics resembles that of a single-field inflation 
modulated by a parameter $\beta$ associated with the coupling 
between the helicity-0 and helicity-1 modes. 

We studied the evolution of longitudinal scalar perturbation $\psi_k$ in the presence of 
the inflaton fluctuation $\delta\phi_k$. The perturbation corresponding to the isocurvature 
mode is given by the combination 
$\delta \chi_k=\psi_k+\beta_m/(2\beta_ A M) \delta \phi_k$. 
Existence of the vector-field mass $M$ comparable to the Hubble expansion rate
during inflation leads to exponential suppression of $\delta \chi_k$ 
after the perturbation enters the region $k^2/a^2<\beta_A M^2$.
We then explicitly showed that the power spectrum of the total curvature perturbation, 
$\mathcal{R}$, generated during inflation, corresponds to that of an effective single-field 
description also corrected by $\beta$. This is possible due to a similar relation between 
$\psi_k$ and $\delta\phi_k$ to that of $A_0$ and $\dot{\phi}$ at the background level, 
obtained in fact by the suppression of $\delta\chi_k$.

After deriving the power spectra of scalar and tensor perturbations generated 
during inflation, we computed their spectral indices $n_s$ and $n_t$ as well as  
the tensor-to-scalar ratio $r$ to confront our inflationary scenario 
with CMB observations. 
The mixing between helicity-0 and helicity-1 modes 
leads to modifications on $n_s$ and $r$ through the parameter 
$\beta$, with the same consistency relation $r=-8n_t$ 
as in the standard canonical case ($\beta=1$).

We computed the observables ${\cal P}_{\cal R}$, $n_s$, and 
$r$ for several inflaton potentials 
to explore the effect of coupling $\beta$ on CMB.
For natural inflation, these observables reduce to those of 
the canonical case after the rescaling of the mass scale $f$. 
In small-field inflation like $\alpha$-attractors and brane inflation, however, 
the coupling $\beta~(<1)$ can lead to the suppression of $r=16\epsilon$
compared to the canonical case. 
This is attributed to the fact that, for smaller $\beta$, the total field velocity 
gets larger and hence inflation needs to start from a region in which 
the potential $V(\phi)$ is flatter to acquire the sufficient amount of 
$e$-foldings. Then, the tensor-to-scalar ratio decreases by the reduction 
of $\epsilon$ on scales relevant to observed CMB anisotropies.

In $\alpha$-attractors given by the potential (\ref{alpo}), we showed that 
$n_s$ and $r$ are approximately given by $n_s \simeq 1-2/N$ and 
$r \simeq 12\alpha_c \beta/N^2$ for $\alpha_c<{\cal O}(10)$. 
This includes the Starobinsky inflation as a special case ($\alpha_c=1$). 
The coupling $\beta$ smaller than 1 leads to the suppression of $r$, so that 
the $\alpha$-attractor model exhibits even better compatibility with 
current CMB observations (see Fig.~\ref{fig1}). 
For $\alpha_c<{\cal O}(10)$, we obtained the observational bound 
$\alpha_c \beta \lesssim 40$ (68 \% CL) from the joint analysis 
based on the \emph{Planck} 2018 $+$ BK14 $+$ BAO data sets. 
The similar suppression of $r$ and the better compatibility with 
observations have been also confirmed for brane inflation 
given by the potential (\ref{Bpo}). 
For $\beta<1$, the brane inflation models with $p=2$ and $p=4$ 
are inside the 95 \% CL and 68~\% CL 
observational contours, respectively,  constrained from the 
\emph{Planck} 2018 $+$ BK14 $+$ BAO data, 
see Fig.~\ref{fig2}.

In this work, we focused on the simple mixing term 
$A^{\mu} \nabla_{\mu}\phi$ as a first step for computing 
primordial power spectra generated during inflation, but the 
further generalization of couplings between $\phi$ and $A^{\mu}$ 
is possible along the lines of Ref.~\cite{Heisenberg2}.
It will be also of interest to study potential signatures of 
such couplings in the CMB bispectrum as well as 
implications in the physics of reheating. Another direct implication
worth studying is the improvement of standard inflationary models 
with respect to the de Sitter Swampland conjecture in the presence 
of this mixing term \cite{Obied:2018sgi}.
These interesting issues are left for future works.

\section*{Acknowledgments}

We would like to thank Jose Beltr\'an Jim\'enez, Claudia de Rham, Ryotaro Kase and Gonzalo Olmo for useful discussions.
HR would like to thank the Institute of Cosmology and Gravitation in Portsmouth for their kind hospitality.
HR\ was supported in part by MINECO Grant SEV-2014-0398,
PROMETEO II/2014/050,
Spanish Grants FPA2014-57816-P and FPA2017-85985-P of the MINECO, and
European Union's Horizon 2020 research and innovation programme under the 
Marie Sk\l{}odowska-Curie grant agreements No.~690575 and 674896.
ST is supported by the Grant-in-Aid 
for Scientific Research Fund of the JSPS No.~16K05359 and 
MEXT KAKENHI Grant-in-Aid for 
Scientific Research on Innovative Areas ``Cosmic Acceleration'' (No.\,15H05890). 

\appendix

\section{ 
Second-order action for scalar perturbations (\ref{Ss2})}
\label{app}

In this Appendix, we show the details for the derivation of 
Eq.~(\ref{Ss2}). In Eq.~(5.4) of Ref.~\cite{HKT18a}, the second-order 
action ${\cal S}_s^{(2)}$ of scalar perturbations was derived 
in general SVT theories by choosing the flat gauge. 
For the specific theories given in this work by Eq.~(\ref{action}), 
we have 
\be
{\cal S}_s^{(2)}=\int dt d^3x\,a^3 \left( 
{\cal L}_s^{\phi}+{\cal L}_s^{\rm GP} \right)\,,
\label{Ssf}
\ee
where
\begin{widetext}
\ba
\hspace{-0.5cm}
{\cal L}_s^{\phi} &=& 
\frac{1}{2} \dot{\delta \phi}^2
-\frac{(\partial \delta \phi)^2}{2a^2}
-\frac{1}{2}V_{,\phi \phi} \delta \phi^2
-\left\{ \dot{\phi} \left( 2-\beta 
\right) \dot{\delta \phi}+V_{,\phi} 
\delta \phi\right\} \alpha
+\dot{\phi}\beta
 \delta \phi \frac{\partial^2 \chi}{a^2}
-\frac{\beta_m M}{2} \left( \dot{\delta \phi} \delta A
-\delta \phi \frac{\partial^2 \psi}{a^2} \right),\\
\hspace{-0.5cm}
{\cal L}_s^{{\rm GP}} &=& 
-2H M_{\rm pl}^2 \alpha \frac{\partial^2 \chi}{a^2}
+\frac{\beta_m^2 \dot{\phi}^2}{2\beta_A^2 M^2 a^2} 
\left[ (\partial \alpha)^2+\frac{\partial^2 \delta A}{A_0}
\alpha+\frac{\partial^2 \dot{\psi}}{A_0} \alpha
+\frac{(\partial \delta A)^2}{4A_0^2}
-\frac{\dot{\psi} \partial^2 \delta A}{2A_0^2}
+\frac{(\partial \dot{\psi})^2}{4A_0^2} \right] \nonumber \\
& &+\left[ \dot{\phi}^2 \left( \frac12+\frac{3\beta_m^2}
{8\beta_A} \right)-3H^2M_{\rm pl}^2 \right]\alpha^2
+\frac{\beta_m^2 \dot{\phi}^2}{8\beta_A} 
\left( \frac{\delta A^2}{A_0^2}-4\alpha \frac{\delta A}{A_0} 
\right)-M^2 \beta_A \frac{(\partial \psi)^2}{2a^2}\,.
\ea
\end{widetext}
Varying the action (\ref{Ssf}) with respect to $\alpha, \chi, \delta A$,
we obtain the three constraint equations in 
Fourier space, respectively as
\begin{widetext}
\ba
& &
\dot{\phi} \left( 1+\frac{\beta_m^2}{4\beta_A}
\right) \dot{\delta \phi}+V_{,\phi} \delta \phi
-\left[ \dot{\phi}^2 \left( 1+\frac{3\beta_m^2}
{4\beta_A} \right)-6H^2 M_{\rm pl}^2 \right]\alpha
+\frac{\beta_m^2 \dot{\phi}^2}{2\beta_A} 
\frac{\delta A}{A_0} \nonumber\\
& &
+\frac{k^2}{a^2} \left[ \frac{\beta_m^2 \dot{\phi}^2}
{2\beta_A^2 M^2} \left( \frac{\dot{\psi}}{A_0}
+\frac{\delta A}{A_0} \right)
-\frac{\beta_m^2 \dot{\phi}^2}{\beta_A^2 M^2} \alpha
-2H M_{\rm pl}^2 \chi \right]=0\,,\label{ap1}\\
& &
\dot{\phi} \left( 1-\frac{\beta_m^2}{4\beta_A}
\right)\delta \phi -2H M_{\rm pl}^2 \alpha=0\,,\\
& &
\beta_m M \dot{\delta \phi}+\frac{\beta_m^2 \dot{\phi}^2}
{2\beta_A} \left( \frac{2\alpha}{A_0}-\frac{\delta A}{A_0^2} 
\right)-
\frac{k^2}{a^2}\frac{1}{A_0} 
\left[ \frac{\beta_m^2 \dot{\phi}^2}
{2\beta_A^2 M^2} \left( \frac{\dot{\psi}}{A_0}
+\frac{\delta A}{A_0} \right)
-\frac{\beta_m^2 \dot{\phi}^2}{\beta_A^2 M^2} \alpha
\right]=0\,.\label{ap3}
\ea
\end{widetext}
We solve Eqs.~(\ref{ap1})-(\ref{ap3}) for $\alpha, \chi, \delta A$ 
and substitute them into Eq.~(\ref{Ssf}). 
Then, in Fourier space, we obtain the second-order action (\ref{Ss2}) 
for dynamical perturbations ${\cal X}^{t}=(\psi_k, \delta \phi_k)$
with the matrix components given by Eq.~(\ref{mass}).



\begin{thebibliography}{99}

\bibitem{Sta80}
A.~A.~Starobinsky,
Phys.\ Lett.\ B {\bf 91}, 99 (1980).

\bibitem{oldinf}
R.~Brout, F.~Englert and E.~Gunzig,
Annals Phys.\  {\bf 115}, 78 (1978);
D.~Kazanas,
Astrophys.\ J.\  {\bf 241} L59 (1980);
K.~Sato, Mon.\ Not.\ R.\ Astron.\ Soc. {\bf 195}, 467 (1981);
Phys.\ Lett.\ {\bf 99B}, 66 (1981);
A.~H.~Guth,
Phys.\ Rev.\ D {\bf 23}, 347 (1981).

\bibitem{oldper}
V.~F.~Mukhanov and G.~V.~Chibisov,
JETP Lett.\  {\bf 33}, 532 (1981);
A.~H.~Guth and S.~Y.~Pi,
Phys.\ Rev.\ Lett.\  {\bf 49} (1982) 1110;
S.~W.~Hawking,
Phys.\ Lett.\ B {\bf 115}, 295 (1982);
A.~A.~Starobinsky,
Phys.\ Lett.\ B {\bf 117} (1982) 175;
J.~M.~Bardeen, P.~J.~Steinhardt and M.~S.~Turner,
Phys.\ Rev.\ D {\bf 28}, 679 (1983).

\bibitem{WMAP} 
G.~Hinshaw {\it et al.} [WMAP Collaboration],
Astrophys.\ J.\ Suppl.\  {\bf 208}, 19 (2013)
[arXiv:1212.5226 [astro-ph.CO]].

\bibitem{Planck2015} 
P.~A.~R.~Ade {\it et al.} [Planck Collaboration],
Astron.\ Astrophys.\  {\bf 594}, A20 (2016)
[arXiv:1502.02114 [astro-ph.CO]].

\bibitem{Planck2018} 
Y.~Akrami {\it et al.} [Planck Collaboration],
arXiv:1807.06211 [astro-ph.CO].

\bibitem{Kolb}
J.~E.~Lidsey, A.~R.~Liddle, E.~W.~Kolb, E.~J.~Copeland,
Rev.\ Mod.\ Phys.\  {\bf 69}, 373 (1997);
D.~H.~Lyth and A.~Riotto,
Phys.\ Rept.\  {\bf 314}, 1 (1999);
B.~A.~Bassett, S.~Tsujikawa and D.~Wands,
Rev.\ Mod.\ Phys.\  {\bf 78}, 537 (2006).

\bibitem{Martin} 
J.~Martin, C.~Ringeval and V.~Vennin,
Phys.\ Dark Univ.\  {\bf 5-6}, 75 (2014)
[arXiv:1303.3787 [astro-ph.CO]].

\bibitem{TOKA} 
S.~Tsujikawa, J.~Ohashi, S.~Kuroyanagi and A.~De Felice,
Phys.\ Rev.\ D {\bf 88}, 023529 (2013)
[arXiv:1305.3044 [astro-ph.CO]].

\bibitem{PTEP} 
S.~Tsujikawa,
PTEP {\bf 2014}, no. 6, 06B104 (2014)
[arXiv:1401.4688 [astro-ph.CO]].

\bibitem{Escudero:2015wba}
M.~Escudero, H.~Ram\'irez, L.~Boubekeur, E.~Giusarma and O.~Mena,
JCAP {\bf 1602}, 020 (2016)
[arXiv:1509.05419 [astro-ph.CO]].

\bibitem{Koi08}
T.~Koivisto and D.~F.~Mota,
JCAP {\bf 0808}, 021 (2008).
[arXiv:0805.4229 [astro-ph]].

\bibitem{Maroto1}
J.~Beltran Jimenez and A.~L.~Maroto,
Phys.\ Rev.\ D {\bf 80}, 063512 (2009).
[arXiv:0905.1245 [astro-ph.CO]].

\bibitem{vecins1}
J.~Beltran Jimenez and A.~L.~Maroto,
JCAP {\bf 0902}, 025 (2009).
[arXiv:0811.0784 [astro-ph]].

\bibitem{vecins2}
G.~Esposito-Farese, C.~Pitrou and J.~P.~Uzan,
Phys.\ Rev.\ D {\bf 81}, 063519 (2010).
[arXiv:0912.0481 [gr-qc]].

\bibitem{vecins3}
P.~Fleury, J.~P.~Beltran Almeida, C.~Pitrou and J.~P.~Uzan,
JCAP {\bf 1411}, 043 (2014).
[arXiv:1406.6254 [hep-th]].

\bibitem{Heisenberg} 
L.~Heisenberg,
JCAP {\bf 1405}, 015 (2014)
[arXiv:1402.7026 [hep-th]].

\bibitem{Allys}
E.~Allys, P.~Peter and Y.~Rodriguez,
JCAP {\bf 1602}, 004 (2016).
[arXiv:1511.03101 [hep-th]].
  
\bibitem{Jimenez16} 
J.~Beltran~Jimenez and L.~Heisenberg,
Phys.\ Lett.\ B {\bf 757}, 405 (2016).
[arXiv:1602.03410 [hep-th]].


\bibitem{BGP} 
G.~Tasinato,
JHEP {\bf 1404}, 067 (2014)
[arXiv:1402.6450 [hep-th]];
G.~Tasinato,
Class.\ Quant.\ Grav.\  {\bf 31}, 225004 (2014)
[arXiv:1404.4883 [hep-th]];
L.~Heisenberg, R.~Kase and S.~Tsujikawa,
Phys.\ Lett.\ B {\bf 760}, 617 (2016)
[arXiv:1605.05565 [hep-th]];
J.~Beltran Jimenez and T.~S.~Koivisto,
Phys.\ Lett.\ B {\bf 756}, 400 (2016)
[arXiv:1509.02476 [gr-qc]];
J.~Beltran Jimenez, L.~Heisenberg and T.~S.~Koivisto,
JCAP {\bf 1604}, 046 (2016)
[arXiv:1602.07287 [hep-th]].

\bibitem{GPcosmo1} 
A.~De Felice, L.~Heisenberg, R.~Kase, S.~Mukohyama, 
S.~Tsujikawa and Y.~l.~Zhang,
JCAP {\bf 1606}, 048 (2016)
[arXiv:1603.05806 [gr-qc]].

\bibitem{GPcosmo2} 
A.~De Felice, L.~Heisenberg, R.~Kase, S.~Mukohyama, 
S.~Tsujikawa and Y.~l.~Zhang,
Phys.\ Rev.\ D {\bf 94}, 044024 (2016)
[arXiv:1605.05066 [gr-qc]].

\bibitem{Bento}
M.~C.~Bento, O.~Bertolami, P.~V.~Moniz, J.~M.~Mourao and P.~M.~Sa,
Class.\ Quant.\ Grav.\  {\bf 10}, 285 (1993)
[gr-qc/9302034].
  
\bibitem{Armen}
C.~Armendariz-Picon,
JCAP {\bf 0407}, 007 (2004)
[astro-ph/0405267].

\bibitem{vectorinf} 
A.~Golovnev, V.~Mukhanov and V.~Vanchurin,
JCAP {\bf 0806}, 009 (2008)
[arXiv:0802.2068 [astro-ph]].

\bibitem{Peloso} 
B.~Himmetoglu, C.~R.~Contaldi and M.~Peloso,
Phys.\ Rev.\ Lett.\  {\bf 102}, 111301 (2009)
[arXiv:0809.2779 [astro-ph]];
B.~Himmetoglu, C.~R.~Contaldi and M.~Peloso,
Phys.\ Rev.\ D {\bf 79}, 063517 (2009)
[arXiv:0812.1231 [astro-ph]].

\bibitem{gaugeinf} 
A.~Maleknejad and M.~M.~Sheikh-Jabbari,
Phys.\ Lett.\ B {\bf 723}, 224 (2013)
[arXiv:1102.1513 [hep-ph]];
A.~Maleknejad and M.~M.~Sheikh-Jabbari,
Phys.\ Rev.\ D {\bf 84}, 043515 (2011)
[arXiv:1102.1932 [hep-ph]].

\bibitem{Soda} 
A.~Maleknejad, M.~M.~Sheikh-Jabbari and J.~Soda,
Phys.\ Rept.\  {\bf 528}, 161 (2013)
[arXiv:1212.2921 [hep-th]].

\bibitem{Namba} 
R.~Namba, E.~Dimastrogiovanni and M.~Peloso,
JCAP {\bf 1311}, 045 (2013)
[arXiv:1308.1366 [astro-ph.CO]].

\bibitem{Ads} 
P.~Adshead, E.~Martinec and M.~Wyman,
JHEP {\bf 1309}, 087 (2013)
[arXiv:1305.2930 [hep-th]].

\bibitem{Davydov}
E.~Davydov and D.~Gal'tsov,
Phys.\ Lett.\ B {\bf 753}, 622 (2016)
[arXiv:1512.02164 [hep-th]]. 

\bibitem{Jose}
J.~Beltran Jimenez, L.~Heisenberg, R.~Kase, R.~Namba and S.~Tsujikawa,
Phys.\ Rev.\ D {\bf 95}, 063533 (2017)
[arXiv:1702.01193 [hep-th]].

\bibitem{aniinf} 
M.~a.~Watanabe, S.~Kanno and J.~Soda,
Phys.\ Rev.\ Lett.\  {\bf 102}, 191302 (2009);
[arXiv:0902.2833 [hep-th]];
A.~E.~Gumrukcuoglu, B.~Himmetoglu and M.~Peloso,
Phys.\ Rev.\ D {\bf 81}, 063528 (2010)
[arXiv:1001.4088 [astro-ph.CO]];
M.~a.~Watanabe, S.~Kanno and J.~Soda,
Prog.\ Theor.\ Phys.\  {\bf 123}, 1041 (2010)
[arXiv:1003.0056 [astro-ph.CO]];
J.~Ohashi, J.~Soda and S.~Tsujikawa,
JCAP {\bf 1312}, 009 (2013)
[arXiv:1308.4488 [astro-ph.CO]].

\bibitem{Turner} 
M.~S.~Turner and L.~M.~Widrow,
Phys.\ Rev.\ D {\bf 37}, 2743 (1988).

\bibitem{Ratra} 
B.~Ratra,
Astrophys.\ J.\  {\bf 391}, L1 (1992).

\bibitem{Bamba} 
K.~Bamba and J.~Yokoyama,
Phys.\ Rev.\ D {\bf 69}, 043507 (2004)
[astro-ph/0310824].

\bibitem{Kanno} 
S.~Kanno, J.~Soda and M.~a.~Watanabe,
JCAP {\bf 0912}, 009 (2009)
[arXiv:0908.3509 [astro-ph.CO]].

\bibitem{Slava} 
V.~Demozzi, V.~Mukhanov and H.~Rubinstein,
JCAP {\bf 0908}, 025 (2009)
[arXiv:0907.1030 [astro-ph.CO]].

\bibitem{Fujita} 
T.~Fujita and S.~Mukohyama,
JCAP {\bf 1210}, 034 (2012)
[arXiv:1205.5031 [astro-ph.CO]].

\bibitem{Mukoh} 
S.~Mukohyama,
Phys.\ Rev.\ D {\bf 94}, 121302 (2016)
[arXiv:1607.07041 [hep-th]].

\bibitem{Heisenberg2} 
L.~Heisenberg,
arXiv:1801.01523 [gr-qc].

\bibitem{Heisenberg:2018vsk} 
L.~Heisenberg,
arXiv:1807.01725 [gr-qc].

\bibitem{Amendola:2012ys} 
L.~Amendola {\it et al.} [Euclid Theory Working Group],
Living Rev.\ Rel.\  {\bf 16}, 6 (2013)
[arXiv:1206.1225 [astro-ph.CO]];
L.~Amendola {\it et al.},
Living Rev.\ Rel.\  {\bf 21}, no. 1, 2 (2018)
[arXiv:1606.00180 [astro-ph.CO]];
E.~J.~Copeland, M.~Sami and S.~Tsujikawa,
Int.\ J.\ Mod.\ Phys.\ D {\bf 15}, 1753 (2006)
[hep-th/0603057];
A.~De Felice and S.~Tsujikawa,
Living Rev.\ Rel.\  {\bf 13}, 3 (2010)
[arXiv:1002.4928 [gr-qc]];
A.~Joyce, B.~Jain, J.~Khoury and M.~Trodden,
Phys.\ Rept.\  {\bf 568}, 1 (2015)
[arXiv:1407.0059 [astro-ph.CO]].

\bibitem{HKT18a} 
L.~Heisenberg, R.~Kase and S.~Tsujikawa,
Phys.\ Rev.\ D {\bf 98}, 024038 (2018)
[arXiv:1805.01066 [gr-qc]].

\bibitem{KT18} 
R.~Kase and S.~Tsujikawa,
JCAP {\bf 1811}, 024 (2018)
[arXiv:1805.11919 [gr-qc]].

\bibitem{HKT18b} 
L.~Heisenberg, R.~Kase and S.~Tsujikawa,
arXiv:1807.07202 [gr-qc] (Physical Review D to appear).

\bibitem{deRham:2012ew} 
  C.~de Rham, G.~Gabadadze, L.~Heisenberg and D.~Pirtskhalava,
  Phys.\ Rev.\ D {\bf 87}, no. 8, 085017 (2013)
  [arXiv:1212.4128 [hep-th]].

\bibitem{natural} 
K.~Freese, J.~A.~Frieman and A.~V.~Olinto,
Phys.\ Rev.\ Lett.\  {\bf 65}, 3233 (1990);
F.~C.~Adams, J.~R.~Bond, K.~Freese, J.~A.~Frieman and A.~V.~Olinto,
Phys.\ Rev.\ D {\bf 47}, 426 (1993)
[hep-ph/9207245].

\bibitem{alpha} 
R.~Kallosh, A.~Linde and D.~Roest,
JHEP {\bf 1311}, 198 (2013)
[arXiv:1311.0472 [hep-th]];
S.~Ferrara, R.~Kallosh, A.~Linde and M.~Porrati,
Phys.\ Rev.\ D {\bf 88}, no. 8, 085038 (2013)
[arXiv:1307.7696 [hep-th]].

\bibitem{Defelice11} 
A.~De Felice, S.~Tsujikawa, J.~Elliston and R.~Tavakol,
JCAP {\bf 1108}, 021 (2011)
[arXiv:1105.4685 [astro-ph.CO]].

\bibitem{branep=2} 
J.~Garcia-Bellido, R.~Rabadan and F.~Zamora,
JHEP {\bf 0201}, 036 (2002)
[hep-th/0112147].

\bibitem{branep=4a} 
G.~R.~Dvali, Q.~Shafi and S.~Solganik,
hep-th/0105203.

\bibitem{branep=4b} 
S.~Kachru, R.~Kallosh, A.~D.~Linde, J.~M.~Maldacena, L.~P.~McAllister and S.~P.~Trivedi,
JCAP {\bf 0310}, 013 (2003)
[hep-th/0308055].

\bibitem{Obied:2018sgi} 
G.~Obied, H.~Ooguri, L.~Spodyneiko and C.~Vafa,
arXiv:1806.08362 [hep-th];
P.~Agrawal, G.~Obied, P.~J.~Steinhardt and C.~Vafa,
Phys.\ Lett.\ B {\bf 784}, 271 (2018)
[arXiv:1806.09718 [hep-th]];
L.~Heisenberg, M.~Bartelmann, R.~Brandenberger and A.~Refregier,
arXiv:1808.02877 [astro-ph.CO];
L.~Heisenberg, M.~Bartelmann, R.~Brandenberger and A.~Refregier,
arXiv:1809.00154 [astro-ph.CO].



\end{thebibliography}
\end{document}